\documentstyle[seceq,epsfig,preprint]{ptptex}

\newcommand{\be}{\begin{equation}}
\newcommand{\ee}{\end{equation}}
\newcommand{\ba}{\begin{array}}
\newcommand{\ea}{\end{array}}
\newcommand{\bc}{\begin{center}}
\newcommand{\ec}{\end{center}}

\newcommand{\prb}[1]{Phys. Rev. {\bf #1}}

\newcommand{\npb}[1]{Nucl. Phys. {\bf #1}}
\newcommand{\la}{\langle}
\newcommand{\ra}{\rangle}
\newcommand{\bpi}{\bar \pi}
\newcommand{\bphi}{\bar \phi}
\newcommand{\bvarphi}{\bar \varphi}

\preprintnumber 
{LPTPE-98-12-01\\ YITP-98-78}
\title{
Mean field theory for collective motion of quantum meson fields
}

\author{
Yasuhiko Tsue\footnote{
Permanent address : Department of Material Science, Kochi University, Kochi 
780-8520, Japan},
Dominique Vautherin and T. Matsui$^\dagger$ 
}

\inst{
Physique Th\'eorique des Particules El\'ementaires, case 127\\ 
Universit\'e Pierre et Marie Curie,
4 place Jussieu, 75252, Paris Cedex 05, France \\
$^\dagger$Yukawa Institute for Theoretical Physics, 
Kyoto University, Kyoto 606-01, Japan \\
}

\recdate{
\today
}

\abst{
Mean field theory for the time evolution of quantum meson fields is 
studied in terms of the functional Schr\"odinger picture with a time-dependent 
Gaussian variational wave functional.  We first show that the equations of 
motion for the variational wavefunctional can be rewritten in a compact
form similar to the Hartree-Bogoliubov equations in quantum many-body theory 
and this result is used to recover the covariance of the theory.  We then 
apply this method to the O(N) model and present analytic solutions of the mean 
field evolution equations for an N-component scalar field.  
These solutions correspond to quantum rotations in isospin space and 
represent generalizations of the classical solutions obtained earlier
by Anselm and Ryskin.  As compared to classical solutions new effects arise
because of the coupling between the average value of the field and its 
quantum fluctuations.  
We show how to generalize these solutions to the case of mean field dynamics 
at finite temperature.
The relevance of these solutions for the observation 
of a coherent collective state or a disoriented chiral condensate in 
ultra-relativistic nuclear collisions is discussed.}

\begin{document}

\maketitle

\section{Introduction}

Several studies have been devoted to the determination of the evolution of a
self-interacting scalar field both at the classical and quantum level.
Some calculations have considered the 
possible formation of a disoriented chiral condensate in ultra-relativistic
nuclear collisions \cite{BLAIZOT,WILCZEK,HOLMAN,SINGH,ANSELM,COOPER}. 
The evolution of a scalar field is also
important to describe some aspects of inflationary models 
\cite{GUTH,JACKIW,DEVEGA,HECTOR,SAMIULLAH,MOTTOLA}.

Most analytical results in this domain concern the classical equations of
motion. This is the case in particular for the rotating solutions in isospin 
space which were investigated some time ago for the non 
linear sigma model by Enikova, Karloukovski and Velchev
\cite{ENIKOVA}, Blaizot and Krzywicki \cite{BLAIZOT} and Anselm 
and Ryskin \cite{ANSELM}. These last authors
discussed the observability of a classical coherent pion field created
during an ultra relativistic nuclear collision in a finite volume. They
examined the effect it would have on the outside volume where it would act
as a source. Anselm and Ryskin 
concluded that a large number of pions, of the order of
fifty in typical collisions, would be emitted in an interaction vertex, all
of them with almost the same momentum. Furthermore some events would
contain mainly neutral pions while some other events would contain in
contrast mainly charged pions.

Classical equations are of interest when the pion density is
high enough to make a description by a coherent state plausible. An
interesting discussion of this limit and several related questions can be
found in the section on waves and particles in the book by Peierls on
``Surprises in Theoretical Physics'' \cite{PEIERLS}. In this section Peierls
points out that physicists originally encountered only the wave
aspects of light and only the corpuscular aspects of particles with mass. He
argues that this is no accident at all because the analogy between light and
matter has very severe limitations. In the case of neutral pions for instance, 
the classical field description implies superpositions of states with different
numbers i.e. corresponding to substantial energy changes. This implies rapid
oscillations in time in the case of free fields. Peierls 
therefore concludes that the question of the
knowledge of the field phase, which is central for a classical description,
becomes of academic interest in this particular case.

In order to have some deeper understanding on the domain of validity of the 
classical field picture and the coherent or collective effects it implies, 
it is highly desirable to have calculational tools incorporating quantum 
effects as well as a consistent treatment of interactions. The most
natural framework for this purpose is the well known mean field or 
Hartree-Bogoliubov approach, which assumes that at each time the density 
matrix of the system is given by the exponential of a quadratic form in 
the field operators (or that the state of the system is described by a 
Gaussian functional in the special case of zero temperature). 

Several numerical investigations of the mean field equations, which generally 
involve large amounts of computational work, are now available for a wide 
variety of geometries and initial conditions. Some analytical results have 
also been obtained concerning the return of the system to equilibrium in 
linear response theory \cite{HECTOR} and the behaviour of mean field solutions 
in a boost invariant geometry at large proper times \cite{PHYREV96}. In our
recent work \cite{PHYLET98} we have used a covariant form of the mean field 
equations to construct a quantum generalization of the time-dependent 
running-wave-type solutions of Anselm and Ryskin for the classical equation 
of motion of non-linear sigma model. 

The purpose of the present paper is three-fold : i) 
First we make a systematic review of the properties of the equations of 
motion in the Gaussian variational approach. 
By introducing a generalized (one-body) density matrix, we show that the 
variational equations of motion can be written in the form of 
Hartree-Bogoliubov like equations, as encountered in many-body theory. 
An attractive property of the density matrix is that its eigenvectors 
are found to be identical to the so-called mode functions. 
These define at each time a set of creation and annihilation operators 
having the Gaussian state as a vacuum. 
The 
mode functions satisfy a set of Klein-Gordon type equation in the presence 
of a mean field 
and they can be exploited to recover the covariance of the theory. 
ii) The second purpose of this paper is to apply the method to the 
O(N) sigma model 
and describe a set of time-dependent solutions which correspond to rotations 
in the internal symmetry space of the quantum fields.  This generalization 
of the classical Anselm-Ryskin solutions was presented in our recent work 
\cite{PHYLET98} also using the covariant formalism.  
iii) In this paper we further show (and this is the main result of this 
present study) that the previous solutions can be extended to finite 
temperature. These solutions allow one to investigate the phase diagram 
of the model as a function of thermal excitation energy 
and/or collective excitation energy.  
  
The present article is organized as follows. We describe the framework of 
the Gaussian approximation in section 2 in the case of a single component 
scalar meson field. The equivalence of this 
approximation with the time dependent Hartree-Bogoliubov
equations is briefly reviewed in section 3. In section 4 a covariant form of
the mean field equations is presented. It uses the so called mode functions,
which are the eigenvectors of the generalized density operator. These
functions have remarkable
properties needed later on to extend the Anselm Ryskin
solutions. In section 5 we consider the case of a scalar field with N
components and an O(N) symmetry. We review
some of the results obtained earlier in this model 
for the classical equations of motion
and describe the structure of the mean field equations. 
We show that rotating solutions in isospin space of the mean field equations 
can be constructed. 
In section 6 an extension to describe mean field equations at finite 
temperature in the O(N) model is given and particular 
analytic solutions of these equations are also presented. They 
correspond to quantum rotations at finite temperature in isospin space.
They include a self consistent treatment of quantum fluctuations around the 
classical solutions of Anselm and Ryskin in the special case of zero momentum 
and zero temperature.
The detailed derivation is included in Appendices A $\sim$ D.
Numerical results are presented in section 7.
We discuss in section 8 the importance of quantum effects, with special focus 
on the self consistency conditions occurring in our equations.
The relevance of our results regarding the possible formation of a coherent 
collective state or a disoriented chiral condensate in ultra relativistic 
collisions is sketched in section 9.

\section{Mean field approximation in the functional Schr\"odinger picture}

We consider in this section a single component
scalar field $\varphi$ described by the Hamiltonian density: 
\be \label{2e1}
{\cal H}({\bf x}) = \frac{1}{2} \pi^2 ({\bf x}) + \frac{1}{2} \left( \nabla 
\varphi ({\bf x}) \right)^2 
+ \frac{m_0^2}{2} \varphi^2 ({\bf x}) + \frac{\lambda}{24} 
\varphi^4 ({\bf x}) \ , \ee
where the operator $\pi ({\bf x})$ is the functional 
derivative $-i \delta/\delta \varphi ({\bf x})$.
The mean field equations are obtained by assuming that at each time $t$ the
wave functional of the system is a Gaussian
\be \label{2e2} 
\Psi \left[ \varphi ({\bf x}) \right] = {\cal N} \exp \left(
i \la  {\bar \pi} | \varphi - {\bar \varphi} \ra 
- \la \varphi- \bar \varphi | ( \frac{1}{4G} + i \Sigma ) | 
\varphi- \bar \varphi  \ra \right)
 , \ee
where $G$, $\Sigma$, $\bar \varphi$, ${\bar \pi}$ 
define respectively the real and imaginary part of the kernel of the
Gaussian and its average position and momentum.
We have used the short hand notation 
\be \label{2e3}
 \la {\bar \pi} |{\varphi} \ra = 
\int {\bar \pi}({\bf x}, t ) \varphi({\bf x}) d{\bf x}.
\ee
The resulting equations are found to be
\be \label{2e4}
\ba{lll}
\displaystyle
\dot G  & = & 2 (G \Sigma +\Sigma G), \\
& \\
\dot \Sigma & = & \frac{1}{8} G^{-2} - 2\Sigma^2 -
\frac{1}{2} \left( -\Delta + m_0^2 + \frac{\lambda}{2} \bar \varphi^2 +
\frac{\lambda}{2}  G({\bf x},{\bf x}) \right) ,
\\
& \\
\dot {\bar \varphi } & = & -\bar \pi ,
\\
& \\
\dot {\bar \pi} & = & 
\left( -\Delta + m_0^2 + \frac{\lambda}{6} \bar \varphi^2 +
\frac{\lambda}{2}  G({\bf x},{\bf x}) \right) \bar \varphi .
\ea
\ee
The vacuum state corresponds to the static solution
\be \label{2e5}
\Sigma=0,~~~~{\bar \pi}=0,~~~~{\bar \varphi}=\varphi_0,~~~G=G_0,
\ee
where
\be \label{2e6}
G_0= \frac{1}{2 \sqrt{-\Delta + \mu^2}}.
\ee
The quantity $\mu$ has to satisfy the so-called gap equation \cite{KERMAN}
\be \label{2e7}
\mu^2= m_0^2 + \frac{\lambda}{2} \la {\bf x}| \frac{1}{2 \sqrt{-\Delta + \mu^2}} |
{\bf x} \ra  + \frac{\lambda}{2} \varphi_0^2 \ .
\ee
In the symmetric phase the expectation value of the field $\varphi_0$ 
vanishes while in the symmetry broken phase it must be such that
\be \label{2e8}
m_0^2 + \frac{\lambda}{2} \la {\bf x}| \frac{1}{2 \sqrt{-\Delta + \mu^2}} |
{\bf x} \ra  + \frac{\lambda}{6} \varphi_0^2=0.
\ee
This last equation implies that $\mu^2$=$\lambda \varphi_0^2/3$.

The previous equation requires a regularization scheme such as a
discretization of the Laplacian operator on a lattice with a mesh size
$\Delta x= 1/\Lambda$ or a cutoff $\Lambda$ in momentum space. To make the
gap equation finite when the scale $\Lambda$ goes to $\infty$ a popular 
prescription is to send the bare coupling constant to zero
according to the formula
\be \label{2e9}
\frac{1}{2 \lambda_R}= \frac{1}{\lambda}+ \frac{1}{16 \pi^2} \log 
(\frac{2 \Lambda}{e \mu}).
\ee
An interesting consequence 
of this prescription is that it also makes the dynamical mean
field equations finite \cite{LIN,SAMIULLAH,MOTTOLA}. Note however that one
difficulty arises because the bare coupling constant becomes infinitesimally
small by negative values when the momentum cutoff goes to infinity, so that
the model is unstable i.e. not really a viable field theory.
This prescription and its limitations have been extensively discussed in the
literature \cite{JACKIW,HOLMAN,KERMAN}. In what follows we will mainly
consider, unless otherwise specified, our interacting scalar field model as
an effective low energy cutoff theory \cite{HOLMAN,ZINN}. The model is
however a useful one which contains attractive aspects. It has in
particular a single bound state whose influence on meson distributions gives
rise to interesting effects \cite{LIN}.
This bound state occurs because the $\lambda \varphi^4$
coupling is a contact force which can accommodate one bound state only
\cite{BEG}.

We have in this section considered only pure states i.e. a zero temperature
theory. The generalization to statistical mixtures and non zero temperatures
is straightforward and the corresponding mean field equations can be found
in the literature \cite{JACKIW}.

\section{Dynamics in Hartree-Bogoliubov form}

It is sometimes convenient to rewrite the mean field evolution equations in
a form which exhibits in a more transparent way the underlying
Lorentz invariance of the equations
and which furthermore reduces to a set of
linear equations for a non interacting meson gas.
Such a form is provided by the familiar time dependent Hartree- Bogoliubov
equations \cite{TSUE,JACKIW,MARTIN} which are particularly suited to
investigate boost invariant configurations. In order to
construct these equations we introduce the
generalized density matrix ${\cal M}$ which adequately implements the
requirements of Lorentz invariance. It is defined by
\be \label{3a1}
{\cal M}({\bf x},{\bf y};t) + \frac{{\bf 1}}{2} = 
\left( \ba{ll} 
i \la {\hat \varphi}({\bf x}) {\hat \pi}({\bf y}) \ra 
& \la {\hat \varphi}({\bf x}) {\hat \varphi}({\bf y}) \ra  \\ 
\la {\hat \pi}({\bf x}) {\hat \pi}({\bf y}) \ra
&-i \la {\hat \pi}({\bf x}) {\hat \varphi}({\bf y}) \ra  
\ea \right),
\ee
where ${\hat \varphi}=\varphi-{\bar \varphi}$ , ${\hat \pi}=\pi-{\bar \pi}$,
${\bf 1}$ is the unit (Dirac) matrix and
expectation values are calculated with the Gaussian functional
$\Psi(t)$. Explicitely
\be \label{3a2}
{\cal M}=
\left( \ba{ll} -2iG \Sigma  & G \\ 
\frac{1}{4G} + 4 \Sigma G \Sigma  & 2i \Sigma G  \ea \right).
\ee
From its structure it can be checked that the generalized density 
matrix satisfies
\be \label{3a3}
{\cal M}^2= \frac{{\bf 1}}{4}.
\ee
The eigenvalues of the density matrix are thus $\pm$ 1/2.
Note that if $(u,v)$ is an eigenvector of ${\cal M}$ with eigenvalue
1/2 i.e.
\be \label{3a4}
\left( \ba{ll} -2iG \Sigma  & G \\ 
\frac{1}{4G} + 4 \Sigma G \Sigma  & 2i \Sigma G  \ea \right)
\left( \ba{l} u \\ v \ea \right)= + \frac{1}{2}
\left( \ba{l} u \\ v \ea \right) ,
\ee
then $(u^*,-v^*)$ is also an eigenvector with eigenvalue -1/2.
Eigenvectors are conveniently normalized to
\be \label{3a5}
\int d {\bf x} \left( u_n^*({\bf x})  v_m({\bf x})+
v_n^*({\bf x})  u_m({\bf x}) \right) = \pm \delta_{m,n},
\ee
with a $\pm$ sign for eigenvalues $\pm 1$/2.
It can also be checked that for eigenvalues +1/2, $u$ and $v$ components
of eigenvectors are related by
\be \label{3a6}
v({\bf x}) = \left( \frac{1}{2 G} + 2i \Sigma \right) u ({\bf x}). 
\ee
The components $u$ of the eigenvectors are often referred to as mode functions
\cite{GUTH,DEVEGA}.

For the particular normalisation condition we have adopted
the eigenvectors $u_n$ and $v_n$ provide the following spectral
decomposition for the generalized density matrix ${\cal M}$
\begin{equation} \label{3a7}
{\cal M}=
\frac{1}{2} \sum_{n>0} \left[ 
\left( \ba{l} u_n \\ v_n \ea \right) \left( v_n^* ~~~u_n^* \right)+
\left( \ba{l} u_n^* \\ -v_n^* \ea \right) \left( -v_n ~~~u_n \right)
\right],
\end{equation}
where the sum runs over eigenstates with eigenvalues +1/2 only.

The interpretation of the mode functions can be made transparent from
the following observation. For a given Gaussian functional $\Psi$ 
specified by its kernel $G$ and $\Sigma$ the operators
\begin{equation} \label{3a8}
b_n~~= \int d{\bf x} \left( v_n({\bf x}) \varphi ({\bf x}) + 
u_n({\bf x}) \frac{ \delta}{ \delta \varphi ({\bf x})} \right) ,
\end{equation}
can be seen to have the state $\Psi$ as a vacuum state for eigenvalues +1/2.
Indeed because of the
relation (\ref{3a6}) between the components $u$'s and $v$'s we have
\begin{equation} \label{3a9}
b_n~|\Psi \ra =0.
\end{equation}
Furthermore because of the normalization condition the operators $b$ and
$b^+$ can be seen to
have canonical commutation relations. The mode functions thus define
a Hartree Bogoliubov transformation preserving commutation relations.

The evolution of the density matrix is governed by the following equations
\cite{MOTTOLA}
\be \label{3a10}
\ba{lll}
\displaystyle
\partial_t G & = & 2 (\Sigma G + G \Sigma), \\
& \\
\partial_t(G^{-1}/4 + 4 \Sigma G \Sigma) & = & 
-2( \Gamma G \Sigma + \Sigma G \Gamma),
\\
& \\
\partial_t \left( \Sigma G \right) & = & 
\frac{1}{8} G^{-1} + 2 \Sigma G \Sigma - \frac{1}{2} \Gamma G,
\\
& \\
\partial_t \left( G \Sigma \right) & = & \frac{1}{8} G^{-1} 
+ 2 \Sigma G \Sigma - \frac{1}{2} G \Gamma .
\ea
\ee
In this equation $\Gamma$ is the mean field
\be \label{3a11}
\Gamma= -\Delta + m_0^2 + \frac{\lambda}{2} \bar \varphi^2 +
\frac{\lambda}{2}  G({\bf x},{\bf x}).
\ee
A remarkable property of the generalized density matrix
is that its equation of motion can be written in Liouville- von
Neumann form
\be \label{3a12}
i {\dot {\cal M}}= [{\cal H}, {\cal M}],
\ee
where the generalized Hamiltonian ${\cal H}$ has the particularly simple form
\be \label{3a13}
{\cal H}= \left( \ba{ll} 0 &1 \\ \Gamma & 0  \ea \right).
\ee

\section{Mean field equations in covariant form}

The equation of motion can also be written in terms of the eigenvectors
$(u,v)$ of the generalized density matrix: 
\be \label{4a1}
i \partial_t \left( \ba{l} u_n \\ v_n \ea \right)=
\left( \ba{ll} 0 &1 \\ \Gamma &0  \ea \right)
\left( \ba{l} u_n \\ v_n \ea \right).
\ee
The normalization condition is preserved by these equations.

Using the explicit form of the mean field operator $\Gamma$ we see that the
mode functions $u_n$ satisfy a set of coupled Klein- Gordon type equations
\be \label{4a2}
\left( \Box + m_0^2 + \frac{\lambda}{2} \bar \varphi^2 +
\frac{\lambda}{2}  G({\bf x},{\bf x}) \right) u_n =0 ,
\ee
while the component $v$ are given by
\be \label{4a3}
v_n= i \partial_t u_n.
\ee
This last equation implies that the normalisation condition can be rewritten
in terms of the mode function $u_n$ only as
\be \label{4a4}
i \int d {\bf x} \left( u_n^*({\bf x})  \partial_t u_m({\bf x})-
 u_m({\bf x}) \partial_t u_n^*({\bf x}) \right) = \pm \delta_{m,n},
\ee
which can be recognized as the familiar normalization of the solutions of
the Klein- Gordon equation \cite{BJORKEN}.

For the self consistent vacuum the mode functions are
\begin{equation} \label{2f17bis}
u_{\bf k} ({\bf x})= 1/\sqrt{2 \omega_{\bf k}} \exp (i {\bf k}. {\bf x}) 
/ (2 \pi) ^{3/2},
\end{equation}
where $\omega_{\bf k}$ is the self consistent energy 
$\sqrt{ {\bf k}^2 + \mu^2}$.

The evolution equations for mode functions
form a closed set. This is because the spectral 
decomposition of
the generalized density matrix ${\cal M}$ implies
\be \label{4a5}
 \la {\bf x}| G(t) |{\bf x} \ra = \frac{1}{2} \sum_n | u_n({\bf x},t)|^2.
\ee
The evolution equations can be written in a more compact and 
manifestly covariant form: 
\be \label{4a6}
m^2(x)= m_0^2 +\frac{\lambda}{2} {\bar \varphi}^2(x) + 
\frac{\lambda}{2} \la x| S |x \ra ,
\ee
where $x= ( x_0, x_1, x_2, x_3 ) $ and where $S$ is the Feynman propagator in
the presence of an $x$ dependent mass
\be \label{4a7}
S= \frac{i}{\Box+m^2(x)+ i \varepsilon}.
\ee
Indeed 
the completeness relation of the eigenvectors of the generalized density
matrix, valid for each time $t$, reads
\be \label{4a8}
\sum_{n>0} \left[ 
\left( \ba{l} u_n \\ v_n \ea \right) \left( v_n^* ~~~u_n^* \right)-
\left( \ba{l} u_n^* \\ -v_n^* \ea \right) \left( -v_n ~~~u_n \right)
\right] = {\bf 1}.
\ee
This relation implies the following ones
\be \label{4a9}
\sum_{n>0} u_n^*({\bf x},t)  i \partial_t u_n({\bf y},t)-
\sum_{n<0} u_n^*({\bf x},t)  i \partial_t u_n({\bf y},t)
= \delta({\bf x}- {\bf y}),
\ee
and
\be \label{4a10}
\sum_{n>0} u_n^*({\bf x},t)  u_n({\bf y},t)-
\sum_{n<0} u_n^*({\bf x},t)  u_n({\bf y},t)=0.
\ee
From these relations we find that the quantity
\be \label{4a11}
-i \sum_{n>0} \left( \theta(t- t') u_n^*({\bf x},t)  u_n({\bf y},t')
+ \theta(t'- t) u_n({\bf x},t)  u_n^*({\bf y},t') \right) ,
\ee
is the inverse of the operator $\partial_{tt} + \Gamma + i \epsilon$, as can
be checked by acting with this operator on the previous quantity. As a result
\be \label{4a12} \ba{lll}
\la x | S | y \ra & = &
\theta(x_0 - y_0 ) \sum_{n>0} u_n^*({\bf x},x_0)  u_n({\bf y},y_0)  \\
&~~\\ 
& & \qquad 
+ ~~ \theta(y_0 - x_0 ) \sum_{n<0} u_n^*({\bf x},x_0)  u_n({\bf y},y_0),
\ea \ee
and
\be \label{4a13}
\la {\bf x}, x_0| S | {\bf x}, x_0 \ra   =  
\la {\bf x}| G(x_0)| {\bf x} \ra.
\ee
This version of the mean field equations is particularly suited to deal with
calculations of the linear response of the meson field to an external source
\cite{PHYREV96}.

At this point let us indicate briefly how the previous equations can be
extended to finite temperatures. The key is that the eigenvalues of the
generalized density matrix are constants of the motion. They are equal to
$\pm$ 1/2 at zero temperature. To have a statistical mixture one has to
consider eigenvalues which differ from this value. For instance one can
consider eigenvalues of the form $f_n$ + 1/2 where the $f$'s are thermal
occupation numbers of bosons.

\section{The case of the O(N) model}

The $O(N)$ model corresponds to the following Lagrangian density
\begin{equation} \label{5a1}
{\cal L}(x)= \frac{1}{2} \partial_{\mu} \phi_a \partial^{\mu} \phi_a
-\frac{1}{2} m_0^2 \phi^2 -\frac{1}{4!} \lambda (\phi^2)^2, 
\end{equation}
where
\begin{equation} \label{5a2}
\phi^2 = \phi_a  \phi_a~~~~a=1,2\ldots N 
\end{equation}
In the special case of the sigma model N=4 the index $a$ corresponds to the
isospin quantum number. In what follows we will be refer to it as
isospin even when N differs from 4.

In the classical approximation the evolution equations read
\begin{equation} \label{5a3}
\left( \Box + m_0^2 + \frac{\lambda}{6} \phi^2(x) \right) \phi_a (x) =0.
\end{equation}

\subsection{The classical solutions of Anselm and Ryskin for the sigma model}

Plane wave like solutions of these equations were obtained 
by Anselm and Ryskin \cite{ANSELM}  who focused on the special case of the
sigma model N=4 and more precisely its non linear version.
Their solutions have the following structure
\begin{equation} \label{5b1}
\phi_a (x) =A_a \cos \left( \omega t -{\bf k}. {\bf r} + \theta_a \right),
\end{equation}
with the conditions
\begin{equation} \label{5b2}
\sum_a A_a^2 \cos(2 \theta_a)= \sum_a A_a^2 \sin(2 \theta_a)=0.
\end{equation}
For these solutions the square of the field is a constant $\varphi_0^2$ 
in space and time
\begin{equation} \label{5b3}
\phi^2 = \varphi_0^2 = \frac{1}{2} \sum_a A_a^2.
\end{equation}
The corresponding energy density is
\begin{equation} \label{5b4}
{\cal E} = \frac{1}{2} \varphi_0^2 ({\bf k}^2+ \omega^2).
\end{equation}
For $\omega$=${\bf k}$=0 (i.e. in the ground state) the constant $\varphi_0^2$
is such that 
\begin{equation} \label{5b5}
m_0^2 + \frac{\lambda}{6} \varphi_0^2=0,
\end{equation}
while for non vanishing momenta it must satisfy
\begin{equation} \label{5b6}
\frac{\lambda}{6} \varphi_0^2= -m_0^2 -{\bf k}^2+ \omega^2.
\end{equation}
We thus see that time like momenta shift the value of the field
outside the chiral radius while the opposite is true for space like
momenta.

For a vanishing value of the momentum the previous solutions depend,
in the case of the sigma model N=4, on six
arbitrary parameters ($A_a, \theta_a, \omega$ with three constraints). For
$A_1^2+ A_2^2 \gg 2 A_3^2$ one has a state with mainly charged pions while
the opposite is true for $A_1^2+ A_2^2 \ll 2 A_3^2$. It was
argued by Anselm and Ryskin \cite{ANSELM} 
that the observation of such events may be a
signature of the formation of a coherent collective state. 

\subsection{Mean field equations at zero temperature}

In the case of the O(N) model the kernels $G$ and $\Sigma$ of the Gaussian
wave functional become N$\times$N matrices while the center $\bphi$
of the Gaussian (often called condensate) 
becomes a vector with N components $\bphi_a$, $a$=1,2\ldots N.
In the mean
field approximation the evolution of ${\bar \varphi}$
is governed by \cite{PHYREV96,CORNWALL}
\begin{equation} \label{MEANFIELD}
\left[ \left( \Box+ m_0^2 + \frac{\lambda}{6} {\bar \varphi}^2 +
\frac{\lambda}{6} {\rm tr} S(x,x) \right) \delta_{ab}+
\frac{\lambda}{3} S_{ab}(x,x) \right] {\bar \varphi}_b (x) =0  \mbox{.}
\end{equation} 
In the previous equation, $S(x,x)$ is a N$\times$N matrix
related to the kernel of the Gaussian
wave-functional
\begin{displaymath}
S_{ab} (x, x) = \langle {\bf x} | G_{ab} (t) | {\bf x} \rangle 
\end{displaymath} 
and the trace runs over the flavour indices.  $S (x, y)$ is the
Feynman propagator 
\begin{equation}\label{PROPAGATOR}
S = \frac{i}{ \Box + m^2(x) - i\epsilon },
\end{equation}
where the N$\times$N mass matrix is  
\begin{eqnarray} \label{MASS}
m_{ab}^2 (x) & = & \left( m_0^2 +\frac{\lambda}{6} {\bar \varphi}^2(x) + 
\frac{\lambda}{6} {\rm tr} S(x,x) \right) \delta_{ab} \nonumber \\
& & \qquad + \frac{\lambda}{3} {\bar \varphi}_a(x) {\bar
\varphi}_b(x) + \frac{\lambda}{3} S_{ab}(x,x) \mbox{.}
\end{eqnarray}
The previous
equations (\ref{MEANFIELD}) -- (\ref{MASS}) are
non-linear because the motion of the condensate involves the mass
matrix $m^2(x)$.
Note that the first three terms in
equation (\ref{MEANFIELD}) correspond to the classical approximation
considered by Anselm and Ryskin.  The next two correspond to the
contribution of quantum fluctuations whose effect is the object of our
study. Generalization of these equations to the case finite temperatures
have been discussed in references 
\cite{BAYM,ROH}.

\subsection{Quantum generalization of the Anselm Ryskin solutions}

Particular solutions of the previous equations which correspond to rotations
in isospin space at zero temperature were already presented in reference
\cite{PHYLET98}. For completeness we give in the present section a short
description of the structure of these solutions. They correspond to the 
following form for the condensate
\begin{displaymath}
{\bar \varphi}(x)= U(x) {\bar \varphi}^{(0)}=
\exp \{i ( q \cdot x ) \tau_y \}
 \left(
\begin{array}{c} 
\varphi_0 \\
0 \\
\vdots \\ 
\end{array}
\right),
\end{displaymath}
where $q_{\mu} =(\omega, {\bf q})$ and $\tau_y $ is a generator of
rotation in the subspace of flavour 1 and 2:
\begin{displaymath}
\tau_y = \left( 
\begin{array}{cccc}
0 & -i & 0 & \cdots \\
i & 0  & 0 & \cdots \\
0 & 0  & 0 & \cdots \\
\vdots & \vdots & \vdots & \ddots 
\end{array}
\right) 
\end{displaymath}
The propagator $S$ is of the form
\begin{displaymath}
S(x,y)= U(x) S^{(0)} (x,y) U^{\dagger}(y),
\end{displaymath} 
with
\begin{displaymath}
S^{(0)} (x,y) = - \int \frac{d^4p}{(2 \pi)^4} S^{(0)} (p) e^{i p \cdot
(x-y)},
\end{displaymath}
and
\begin{displaymath}
S^{(0)} (p)= \frac{i}
{(p_{\mu} + q_{\mu} \tau_y)(p^{\mu} + q^{\mu} \tau_y)
-M^2 + i \varepsilon}.
\end{displaymath}
It can be seen that the previous expressions 
solve the mean field equations provided the mass matrix $M$
in the propagator $S^{(0)}$ satisfies
\begin{eqnarray}
M_{ab}^2 &=& \left(m_0^2 +\frac{\lambda}{6} \varphi_0^2 
+ \frac{\lambda}{6} {\rm tr} S^{(0)} (x,x) \right) \delta_{ab} 
\nonumber\\
& & \qquad \qquad \qquad
+ \frac{\lambda}{3} {\bar \varphi}_a^{(0)} {\bar \varphi}_b^{(0)} 
+ \frac{\lambda}{3} S_{ab}^{(0)} (x,x) \mbox{.}
\label{gap}
\end{eqnarray}
To have a closed set, the previous equations must be supplemented by
the relation satisfied by the condensate ${\bar \varphi}^{(0)}$
\begin{eqnarray}
& & \left[ -q^2 + m_0^2 + \frac{\lambda}{6} \varphi_0 ^2 +
\frac{\lambda}{6} {\rm tr} S^{(0)} (x,x) 
+ \frac{\lambda}{3} S^{(0)}_{11}(x,x) \right] \varphi_0 =0 \mbox{.} 
\label{condensate}
\end{eqnarray}
The above solutions depend on the 4 parameters $(\omega, {\bf q})$. For
fixed values of these parameters the above equations provide the values of
the matrix elements $M_{ab}$ and of the condensate $\varphi_0$. The matrix
elements $M_{ab}$ are found to be diagonal in the isospin index i.e. 
$M_{ab}$=
$M_a \delta_{ab}$. For rotations in the subspace of flavour 1 and 2 one has
$M_3$= $M_4$= $\mu$ and one thus needs only to determine the four quantities 
$M_1$, $M_2$, $\mu$ and $\varphi_0$.  
In reference \cite{PHYLET98} it was shown, using a perturbative analysis,
that a disappearance of the condensate $\varphi_0$, reflecting
a restauration of chiral symmetry, occurs for $\omega$=0 at values of ${\bf q}$
corresponding to an energy density of about (123 MeV)$^4$. We refer the
reader to reference \cite{PHYLET98} for a discussion of this result. In the
next section we focus on an extension of this calculation to the case of
rotating solutions in isospin space at finite temperature. 

\section{Mean field equations at finite temperature in the O(N) model}

The mean field equations at finite temperature in the O(N) model are
presented and discussed in Appendix A. The construction of rotating
solutions in isospin space at finite temperature is presented in
Appendix B.  These solutions are constructed in two successive
steps. The first step is to write the mean field evolution equations
in a rotating frame in isospin space. The second one is to look for
static solutions in the rotating frame.  In the present section we
will just give a short description of the structure of the solutions.
As compared to the previous section, devoted to the zero temperature
case, the basic structure of the coupled gap equations (\ref{gap}) and the
condensate equation (\ref{condensate}) is unchanged.  The only
difference is that the propagator in the rotating frame now involves
boson occupation numbers: 
\begin{equation} \label{6d2}
S(p) =(2n(p_0)+1) 
\frac{i}{(p_0-\omega\tau_y)^2-({\bf p}-{\bf q}\tau_y)^2-M^2+i\epsilon} \ ,
\end{equation}
where
\begin{equation} \label{6d3}
n(p_0) = \frac{1}{{\rm e}^{\beta p_0}-1},
\label{occup}
\end{equation}
as shown in Appendix C, see Eqs.(\ref{C23}) and (\ref{C19}).  Here,
$\omega$ is the frequency of isospin rotation and ${\bf q}$ is its
wave number.  The appearance of the occupation number (\ref{occup})
breaks the Lorentz invariance of our previous results for the vacuum
case. The non-zero value ${\bf q}$ means the space-oscillation is
realized in isospin rotating frame where the condensate is static and
uniform.  The above expression contains a specific information of the
boundary conditions by the $\epsilon$-prescription for the solution of
a generalized Klein-Gordon equation as shown in detail in Appendix D
using the mode-expansion of the propagator, (\ref{D3}).  It is shown
that the gap equations can be written in the form as in the vacuum
(\ref{gap}) with fluctuation terms involving the matrix: 
\begin{equation}\label{6a1}
S_{ab}(x,x)= \int \frac{d^3{\bf p}}{(2\pi)^3} 
S_{ab}({\bf p}; x_0=0)\ , 
\end{equation}
where
\begin{equation}\label{6d4}
S({\bf p}; x_0=0 )
=-\int\frac{d p_0}{2\pi i} (2n(p_0)+1)
\frac{1}{(p_0-\omega\tau_y)^2-({\bf p}-{\bf q}\tau_y)^2-M^2+i\epsilon} \ .
\end{equation}
The integral with respect to $p_0$ can be carried out analytically: it
picks up the poles of the integrand in the lower complex $p_0$-plane
as in the vacuum case.  We present below the results for the pure
time-like rotation ( $q^2=\omega^2  > 0$ , ${\bf q} =0$) and for the
pure space-like rotation ($q^2= -{\bf q}^2  <  0$ , $\omega =0$ ). For
both cases, off-diagonal matrix elements of $S_{ab}( {\bf p}, x_0 = 0 )
$ vanish after integration over $p_0$ or $\bf p$.  This implies that
the matrix $M_{ab}^2$ is diagonal as in the case for the vacuum.

The gap equations can therefore be written explicitly as
\begin{eqnarray}
M_1^2 & = & m_0^2 + \frac{\lambda}{2} \varphi_0^2 
+ \frac{\lambda}{6} \sum_{c=1}^N S_{cc} (x,x) 
+ \frac{\lambda}{3} S_{11} (x,x) \nonumber\\
M_2^2 & = & m_0^2 + \frac{\lambda}{6} \varphi_0^2 
+ \frac{\lambda}{6} \sum_{c=1}^N S_{cc} (x,x) 
+ \frac{\lambda}{3} S_{22} (x,x) \nonumber\\
\mu^2 & \equiv & M_3^2 = M_4^2 = \cdots = M_N^2 \nonumber\\
& = & m_0^2 + \frac{\lambda}{6} \varphi_0^2 
+ \frac{\lambda}{6} \sum_{c=1}^N S_{cc} (x,x) 
+ \frac{\lambda}{3} S_{33} (x,x) \ , \nonumber\\
\end{eqnarray}
while the condensate equation becomes
\begin{equation} \label{6d10}
\frac{\lambda}{6} \varphi_0^2
=\omega^2-{\bf q}^2-m_0^2-\frac{\lambda}{6}\sum_{c=1}^{N}
S_{cc}(x,x) -\frac{\lambda}{3} S_{11}(x,x) \ . 
\end{equation} 
By using the mass $M_1^2$, the above condensate equation can be recast
into a simpler form: 
\begin{equation} \label{6d11}
\frac{\lambda}{3}\varphi_0^2 = -\omega^2+{\bf q}^2+M_1^2 \ .
\end{equation} 

The explicit forms of the integrands of the diagonal matrix elements
of $S_{ab}({\bf p}, x_0=0)$ are given below for two specific cases.

\vskip 0.7cm
\leftline{{\bf Case 1)} Pure time-like isospin rotations ($ {\bf q} = 0 $): }
\vskip 0.5cm

\begin{eqnarray} \label{6d5}
S_{11}({\bf p}, x_0=0)&=&  
(2n(E_+)+1)\left[\frac{1}{4E_+}+\frac{4\omega^2+M_1^2-M_2^2}
{4E_+(E_+^2-E_-^2)}
\right] \nonumber\\
& &\qquad\qquad
+(2n(E_-)+1)\left[\frac{1}{4E_-}-\frac{4\omega^2+M_1^2-M_2^2}
{4E_-(E_+^2-E_-^2)}
\right] \ , \nonumber\\
S_{22}({\bf p}, x_0=0)&=&
(2n(E_+)+1)\left[\frac{1}{4E_+}+\frac{4\omega^2+M_2^2-M_1^2}
{4E_+(E_+^2-E_-^2)}
\right] \nonumber\\
& &\qquad\qquad
+(2n(E_-)+1)\left[\frac{1}{4E_-}-\frac{4\omega^2+M_2^2-M_1^2}
{4E_-(E_+^2-E_-^2)}
\right]  \ , 
\end{eqnarray}
where
\begin{equation} \label{6d6}
E_{\pm}^2({\bf p})= {\bf p}^2+\frac{M_1^2+M_2^2}{2}+\omega^2\pm
\sqrt{\left(\frac{M_1^2-M_2^2}{2}\right)^2+4\omega^2
\left({\bf p}^2+\frac{M_1^2+M_2^2}{2}\right)} \ .
\end{equation}
We note that $E_{\pm}$ are also eigenvalues for ${\cal H}(\omega)$ in
(\ref{B3}).  Also, the matrix elements $S_{ab}(x,x)$ for $a, b\ge 3$
are given as
\begin{equation} \label{6d7}
S_{ab}(x,x)
= \delta_{ab} \int \frac{d^3{\bf p}}{(2\pi)^3} 
\frac{1}{\sqrt{{\bf p}^2+\mu^2}}\left(
n(\sqrt{{\bf p}^2+\mu^2})+\frac{1}{2} \right) \qquad 
({\rm for}\ a,b=3,4,\cdots, N) \ .
\end{equation}

\vskip 0.7cm
\leftline{{\bf Case 2)} Pure space-like isospin rotations ($\omega =0$): }
\vskip 0.5cm

The expressions of the diagonal matrix elements, $S_{11}(x,x)$ and
$S_{22}(x,x)$ are slightly different from those derived in the
time-like case, while those of $S_{cc}(x,x)$ for $c \geq 3$ are
unmodified:
\begin{eqnarray} \label{6d8}
S_{11}({\bf p}, x_0=0)
&=&
(2n(E_+)+1)\left[\frac{1}{4E_+}+\frac{M_1^2-M_2^2}{4E_+(E_+^2-E_-^2)}
\right] \nonumber\\
& &\qquad\qquad
+(2n(E_-)+1)\left[\frac{1}{4E_-}-\frac{M_1^2-M_2^2}{4E_-(E_+^2-E_-^2)}
\right] \ , \nonumber\\
S_{22}({\bf p}, x_0=0)
&=&
(2n(E_+)+1)\left[\frac{1}{4E_+}-\frac{M_1^2-M_2^2}{4E_+(E_+^2-E_-^2)}
\right] \nonumber\\
& &\qquad\qquad
+(2n(E_-)+1)\left[\frac{1}{4E_-}+\frac{M_1^2-M_2^2}{4E_-(E_+^2-E_-^2)}
\right] \ , 
\end{eqnarray}
where
\begin{equation}\label{6d9}
E_{\pm}^2({\bf p})
={\bf p}^2+{\bf q}^2+\frac{M_1^2+M_2^2}{2} \pm
\sqrt{\left(\frac{M_1^2-M_2^2}{2}\right)^2+4({\bf p}\cdot {\bf q})^2} \ .
\end{equation}

\section{Numerical results}

We can numerically solve the gap equation (\ref{6d10}) with 
(\ref{6a1}), (\ref{6d5}) 
and (\ref{6d7}) for $q^2=\omega^2 \geq 0$, or (\ref{6a1}), 
(\ref{6d8}) and (\ref{6d7}) 
for $q^2=-|{\bf q}|^2 < 0$ in a self-consistent manner. 
We set the number of isospin $N=4$. A three-dimensional momentum cutoff 
$\Lambda = 1$ GeV is introduced. 
For this cutoff, the model parameters used here, $\lambda$ and $m_0^2$, 
are taken so as to reproduce the chiral condensate $\varphi_0$ to be 
the pion decay constant $f_{\pi}=93$ MeV and $M_1$ to be the sigma meson mass 
$M_{\sigma}=500$ MeV at $q_\mu = 0$ and $T=0$. 
Then, condensate equation at $T=q_\mu = 0$ gives $\lambda=86.7$ and 
$m_0^2 =-(1019.5 {\rm MeV})^2$. 
One-dimensional numerical integration for $|{\bf p}|$ in Eq.(\ref{6d5}) 
is performed for $q^2=\omega^2 \geq 0$ case. 
For $q^2=-|{\bf q}|^2 < 0$, two-dimensional integration is required because 
the square of the inner product $({\bf p}\cdot {\bf q})^2$ appears in 
(\ref{6d8}) with (\ref{6d9}). 
However, we replace $\cos ^2 \theta$ with its angle-averaged value 
$\la \cos ^2 \theta \ra = 1/3$. 
We safely perform one-dimensional integration 
for $|{\bf p}|$ with the above-mentioned replacement 
because the resultant curves for condensate $\varphi_0$ 
versus $|{\bf q}|$ are indistinguishable at $T=0$.
We checked numerically that there is no visible difference at $T=0$. 
\begin{center}
\begin{figure}
\vspace{10cm}
\includegraphics{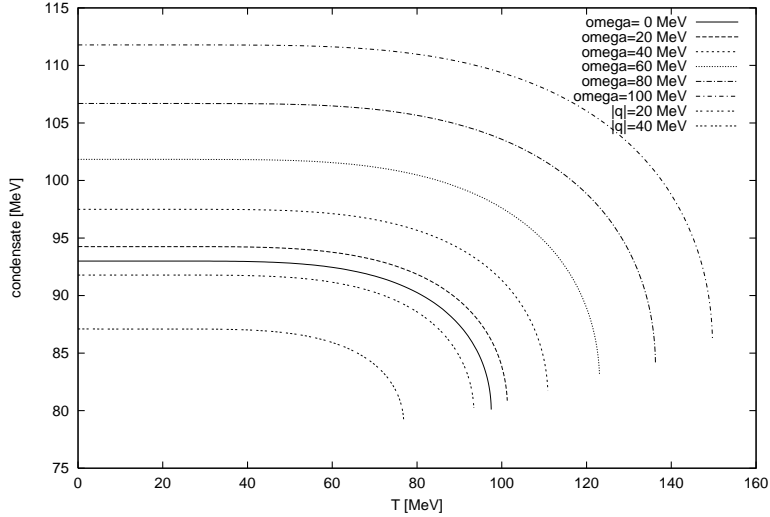}
\caption{The chiral condensate $\varphi_0$ with $q^2$ is depicted as 
a function of temperature. The vertical axis represents the chiral condensate 
$\varphi_0$ (MeV) and the horizontal axis is temperature (MeV). 
The solid curve corresponds to the case $q_{\mu}=0$. 
Upside five curves from the solid curve correspond to the cases  
$\sqrt{q^2}=\omega=$20, 40, 60, 80 and 100 MeV, and downside 
two curves from the solid curve correspond to the cases 
$\sqrt{-q^2}=|{\bf q}|=$
20 and 40 MeV. 
}
\label{Fig1}
\end{figure}
\end{center}

The chiral order parameter with a certain $q^2$ is shown as a function of 
temperature in Fig. 1. 
The increase of temperature leads to the decrease of the amplitude of 
the chiral order parameter $\varphi_0$ at each $q^2$. 
Namely, chiral symmetry is partially restored at finite temperature. 
At a certain critical temperature $T_c$, a solution of the gap equation does 
not exist suddenly. 
At $T > T_c$, only trivial solution $\varphi_0=0$ exists. 
The order of phase transition is like the first order. 
In our numerical calculation, $T_c$'s are 
97.5, 101.3, 110.9, 123.1, 136.3 and 149.7 MeV for 
$\sqrt{q^2}=\omega=$0, 20, 40, 60, 80 and 100 MeV, respectively, and 
93.4 and 76.9 MeV for $\sqrt{-q^2}=|{\bf q}|=$ 20 and 40 MeV, respectively.

Let us consider the case of time-like $q^2$, that is,
$q^2=\omega^2 \geq 0$ with $|{\bf q}|=0$. 
As $\omega$ increases at each $T < T_c$, the condensate $\varphi_0$ also 
increases. In this case the condensate rotates uniformly with angular 
frequency $\omega$ in isospin space. 
As a result, the chiral symmetry breaking is enhanced. 
\begin{center}
\begin{figure}
\vspace{10cm}
\includegraphics{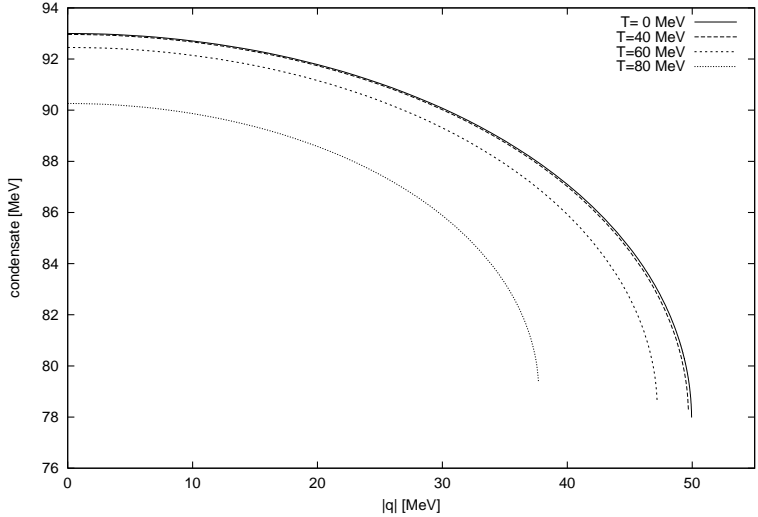}
\caption{The chiral condensate $\varphi_0$ with temperature $T$ 
is depicted as 
a function of three-momentum $|{\bf q}|$. 
The vertical axis represents the chiral condensate $\varphi_0$ (MeV) 
and the horizontal axis 
represents the magnitude of three momentum $|{\bf q}|$ (MeV). 
The solid curve corresponds to 
the zero temperature case. 
Three curves from the solid curve correspond to the cases  
$T=$40, 60 and 80 MeV. 
}
\label{Fig2}
\end{figure}
\end{center}

In the case of space-like $q^2$, that is, $q^2=-|{\bf q}|^2 < 0$ with 
$\omega=0$, the increasing $|{\bf q}|$ results in the decreasing $\varphi_0$ 
at each $T<T_c$. 
Further, the critical momentum $|{\bf q}_c|$ exists as is shown in Fig. 2. 
Namely, the gap equation becomes to have no non-trivial solution 
at a certain value $|{\bf q}_c|$. 
The transition seems the first order phase transition. 
For example, 
$|{\bf q}_c|$'s are 50.0, 49.9, 49.7, 47.2 and 37.7 MeV for $T=$ 0, 20, 40, 
60 and 80 MeV, respectively.

It is interesting to compare these critical momenta for space-like condensate 
to the corresponding values in the classical limit. 
In classical case, $|{\bf q}_c|^2=M_{\sigma}^2/2$ is expected at $T=0$ 
from the classical gap equation. However, the quantum effect leads to more 
rapid change of chiral condensate as was pointed out by perturbation theory 
in the previous paper.\cite{PHYLET98}
This is expected because quantum fluctuation smears out the effective 
potential and makes symmetry breaking more difficult to reach. 
The coupling to the quantum fluctuation works to suppress the appearance of 
static condensate with longer wavelength. This would have important 
implications for the dynamics of chiral condensate in high energy 
nuclear collisions.

\section{Importance of quantum effects}

In the work of Anselm and Ryskin, which uses the framework of the classical 
equations of motion, it was pointed out that the existence of rotating 
solutions 
in isospin space may have observable consequences in ultra-relativistic 
collisions.
For instance it may be possible to produce a classical state of the pion 
field in 
which all pions nearly carry the same momentum, which would manifest 
itself by the observation of pion ``jets''. 
Since Anselm and Ryskin's analysis was restricted to the classical picture, 
an important question is to find out how the previous scenario is 
affected by quantum fluctuations.
A possible way to evaluate the importance of these effects is to compare
the rotational contributions to the energy density arising from classical and
quantum terms. The energy density at a given momentum is indeed 
proportional to the pion number 
density times the pion energy.
In the present framework the rotational
contribution ${\cal E}_R$ to the energy density (due to non-vanishing
$\Sigma$ and $\bpi$ ) at zero temperature is
\begin{equation} \label{8a1}
{\cal E}_R({\bf x}) = \frac{1}{2} {\bar \pi}^2 + 2~ {\rm Trace} 
\la {\bf x}|\Sigma G \Sigma|{\bf x} \ra .
\end{equation}
(See, (\ref{A3}) with $\xi=\sigma=0$ at zero temperature in Appendix A.)
The first term is of classical origin while the second term is purely
quantal. 
Using the fact that 
$G_{ab}=S_{ab}$ with $n(E)=0$ at zero temperature and 
by solving the equation of the second line in (\ref{A11}) to get $\Sigma$ 
with $i{\dot G}^{-1}$ being $i\omega [G^{-1}, \tau_y]$ in isospin rotating 
frame,
this contribution can
be rewritten as
\begin{equation} \label{8a2}
{\cal E}_R({\bf x})= \frac{1}{2} \omega^2 \left( \varphi_0^2 + I \right),
\end{equation}
with
\begin{equation} \label{8a3}
I=  \int \frac{d {\bf p}}{(2 \pi)^3}~
\frac{\left( S_{22}({\bf p}, x_0=0)-S_{11}({\bf p}, x_0=0) \right)^2}
     {S_{22}({\bf p}, x_0=0)+S_{11}({\bf p}, x_0=0)}.
\end{equation}
Let us evaluate this expression in the vicinity of $\omega \simeq 0$ 
at $T=0$. In
this case we replace the kernels $S$ by $1/2\sqrt{p^2 + M^2}$ with the
result
\begin{equation} \label{8a4}
I \simeq  \mu^4 \int \frac{d {\bf p}}{(2 \pi)^3}~
\frac{ 1 }{ 
\sqrt{\left( p^2 + M_1^2 \right) \left( p^2 + M_2^2 \right) } 
\left( \sqrt{p^2 + M_1^2} + \sqrt{p^2 + M_2^2} \right)^3 } , 
\end{equation}
where $M_1^2 = M^2 + \mu^2$ and $M_2^2 = M^2 - \mu^2$. 
As a first approximation we replace the square masses in the denominator 
by their averages $M^2$ so that
\begin{equation} \label{8a5}
I \simeq  \frac{1}{4} \mu^4 \int \frac{d {\bf p}}{(2 \pi)^3}~
\frac{1}{\left( p^2 + M^2 \right)^{5/2}}
= \frac{\mu^4}{ 24 \pi^2 M^2}. 
\end{equation}
This gives the approximate contribution to the rotational energy density
\begin{equation} \label{8a6}
{\cal E}_R({\bf x}) \simeq \frac{1}{2} \omega^2 \left( \varphi_0^2 + 
\frac{\mu^4}{ 24 \pi^2 M^2} \right).
\end{equation}
Taking the value of the condensate $\varphi_0$ to be the pion decay
constant $f_{\pi} = 93 $ MeV, $M_1$ to be the sigma meson mass
$M_{\sigma} = 500 $ MeV and $M_2$ to be of the order of the pion mass
$M_{\pi} = 140 $ MeV we find that quantum effects give only a small
contribution ( $4.8$ \%) to the energy density.
We can also evaluate the quantum contribution (\ref{8a3}) directly 
in our numerical calculation with three-momentum cutoff. 
We conclude that the quantum effects yield only a very small contribution 
about $2.8$ \% in the range of $\omega=0 \sim 100$ MeV.

We thus expect the number of pions outside of the condensate to be small. 
This implies that, in this case, the 
discussion of Anselm and Ryskin is unaffected 
by quantum fluctuations. Let us recall however that quantum fluctuations still 
play an important role since they drive the chiral radius away from its vacuum 
value as excitation energy increases.

\section{Discussion}

In the present paper we have shown that it is possible to construct
analytically quantum coherent states which are
solutions of the mean field evolution equations 
in the case of an N-component scalar field. 
We have proceeded in two steps. We have first
constructed rotating solutions in isospin space in the zero temperature case. 
In a second step we have generalized our results to the case of finite 
temperature. 
These solutions represent the quantum generalization 
of the classical solutions obtained earlier for the sigma model
by Anselm and Ryskin \cite{ANSELM}.

As compared to the work of Anselm and Ryskin we have found that new effects 
arise as a result of the coupling between the motion of the mean (classical) 
value of the field and its quantum fluctuations. In particular restauration 
of chiral symmetry is affected by quantum fluctuations. 
Another difference is that, while classical solutions
correspond to a distribution of mesons with almost the same momentum, a new
component containing all possible momenta appears in the quantum solution. 
However we have found that the number of mesons outside of the condensate 
is small. As a result the conclusion of Anselm and Ryskin about the 
observability of the coherent classical state remains unaffected.

Further questions in the line of the study by Anselm and Ryskin are worth
additional investigations. For instance it would be interesting to consider an
initial configuration in which the coherent solution we have built is
present in a finite volume. By considering this field as a source for the
external volume it would be possible to find out the distribution of the
emitted pions as discussed by Anselm and Ryskin \cite{ANSELM}.

\section*{Acknowledgements}

One of us (D. V.) wishes to thank Prof. Yosuke Nagaoka and the members of the
Yukawa Institute for Theoretical Physics of the Kyoto University for the
hospitality extended to his visit. He also wishes to express his
appreciation to the Japan Society for the Promotion of Science for support
to visit the Yukawa Institute.  T. M. is grateful to the members of
Division de Physique Th\'eorique, Institut de Physique Nucl\'eaire,
Orsay, for hospitality during his visit. 
Y. T. would like to express his thanks to the members of the 
Laboratoire de Physique Th\'eorique des Particles El\'ementaires 
(LPTPE) of Universit\'e Pierre et Marie Curie for the warm hospitality. 
He also wishes to express his sincere thanks to Nishina Memorial Foundation 
for financial support to visit the LPTPE.

%
%

\appendix

\section{Mean field equations of motion at finite temperature}

Let us consider in this Appendix the quantum and thermal generalization 
of the classical equations of the O(N) model. 
Following the Eboli, Jackiw and Pi prescription,\cite{JACKIW}
let us introduce a Gaussian density 
matrix in the functional Schr\"odinger picture : 
\begin{eqnarray} \label{A1}
\rho[\varphi_1, \varphi_2]&=&{\cal N}_G
\exp\biggl(-i\la {\bar \pi}|\varphi_1-\varphi_2 \ra
-\la \varphi_1 - {\bar \varphi}|(\frac{1}{4G}-i\Sigma)|\varphi_1-{\bar \varphi}
\ra \nonumber\\
& &\qquad\quad
-\la \varphi_2 - {\bar \varphi}|(\frac{1}{4G}+i\Sigma)|\varphi_2-{\bar \varphi}
\ra
+\frac{1}{2}\la \varphi_1 - {\bar \varphi}|
\frac{1}{\sqrt{G}}\zeta\frac{1}{\sqrt{G}}|\varphi_2-{\bar \varphi}
\ra \biggl)\ , \nonumber\\
& & 
\end{eqnarray}
where ${\cal N}_G$ is a normalization factor. Here $\zeta$ is a mixing 
parameter and we divide it into real and imaginary parts as
$\zeta = \xi + 4i\sqrt{G}\sigma\sqrt{G}$.The real part, $\xi$, is a 
symmetric kernel ($\xi({\bf x}, {\bf y})=\xi({\bf y},{\bf x})$) and 
the imaginary part $\sigma$ is an antisymmetric one. 
If $\zeta=0$, the Gaussian density matrix is expressed in terms of 
pure states by 
$\rho[\varphi_1, \varphi_2]=\Psi[\varphi_2] \Psi^*[\varphi_1]$.
Averaged values are easily calculated by
\begin{equation}\label{A2}
\la {\cal O} \ra = {\rm Trace}(\rho {\cal O})
=\int\int{\cal D}\varphi_1{\cal D}\varphi_2\ 
\rho[\varphi_1, \varphi_2] {\cal O}[\varphi_2, \varphi_1] \ , 
\end{equation}
where ${\cal O}[\varphi_2, \varphi_1] \equiv \la \varphi_2|{\cal O}|\varphi_1
\ra$.

The energy density calculated for such a Gaussian density matrix is found to be
\begin{equation} \label{A3}
\begin{array}{lll}
\displaystyle
\la {\cal H}({\bf x}) \ra  & = & {\cal E}_0({\bf x}) + 
\frac{1}{8} {\rm Trace} \la {\bf x}|G^{-\frac{1}{2}}(1+\xi)
G^{-\frac{1}{2}}|{\bf x} \ra \\
& \\
& & \qquad 
+ \frac{1}{2} {\rm Trace} \la {\bf x}|( - \Delta + m_0^2) 
\left(G^{\frac{1}{2}}\frac{1}{1-\xi}G^{\frac{1}{2}}\right)|{\bf x} \ra  \\
& \\
& & \qquad + ~ 2 ~ {\rm Trace} \la {\bf x}| (\Sigma+\sigma) 
G^{\frac{1}{2}}\frac{1}{1-\xi}G^{\frac{1}{2}}( \Sigma -\sigma)|{\bf x} \ra + 
\la V({\bf x}) \ra ,
\end{array}
\end{equation}
where 
\begin{equation} \label{A4}
{\cal E}_0 = \frac{1}{2} {\bar \pi}_a^2 + \frac{1}{2} ( \nabla \bvarphi_a )^2 
+ \frac{1}{2} m_0^2 \bvarphi^2 + \frac{1}{24} \lambda ( \bvarphi^2 )^2 
\end{equation}
is the classical energy density and the Trace 
is to be taken over the isospin index $a$.
The last term in this equation is expressed as 
\begin{equation} \label{A5}
\begin{array}{lll}
\displaystyle
& &{\hbox{\hspace{-0.8cm}}}
 \la V({\bf x}) \ra = \frac{1}{24} \lambda \left(
2  \bvarphi^2 \la {\hat \varphi}_a ({\bf x}){\hat \varphi}_a ({\bf x}) \ra + 
\la {\hat \varphi}_a ({\bf x}){\hat \varphi}_a ({\bf x}) \ra
\la {\hat \varphi}_b ({\bf x}){\hat \varphi}_b ({\bf x}) \ra \right. \\
& \\
& & \qquad \left.+ 4 \bvarphi_a({\bf x})
\la {\hat \varphi}_a ({\bf x}){\hat \varphi}_b ({\bf x}) \ra 
\bvarphi_b({\bf x}) 
+ 2 \la {\hat \varphi}_a ({\bf x}){\hat \varphi}_b ({\bf x}) \ra 
\la {\hat \varphi}_a ({\bf x}){\hat \varphi}_a ({\bf x}) \ra
\right) , 
\end{array}
\end{equation}
where 
$\la {\hat \varphi}_a ({\bf x}){\hat \varphi}_b ({\bf x}) \ra 
\equiv [G^{1/2}(1-\xi)^{-1}G^{1/2}]_{ab}({\bf x}, {\bf x})$ and 
repeated indices are supposed to be summed over.

The classical equations of motion for $\bvarphi_a({\bf x}, t)$ and 
$\bpi_a({\bf x}, t)$ in quantum and thermal fluctuations are written as 
\be \label{A6}
\ba{lll}
\displaystyle
& &{\hbox{}\hspace{-0.6cm}}
\dot{\bvarphi}_a({\bf x}, t) =  -{\bpi}_a({\bf x}, t) ,
\\ && \\
& &{\hbox{}\hspace{-0.6cm}}
\dot{\bpi}_a({\bf x}, t) =  
\left( -\Delta + m_0^2 + \frac{\lambda}{6} \bvarphi^2 +  \frac{\lambda}{6} 
 \la {\hat \varphi}_c({\bf x}){\hat \varphi}_c({\bf x})\ra
 \right) \bvarphi_a 
+ \frac{\lambda}{3} \la {\hat \varphi}_a({\bf x}){\hat \varphi}_b({\bf x})\ra
 \bvarphi_b.
\ea
\ee

Let us introduce the generalized density matrix in O(N) model as is similar to 
that of $\phi^4$-theory : 
\begin{eqnarray} \label{A7}
{\cal M}^{ab}({\bf x},{\bf y};t) &=& \la {\bf x}|{\cal M}^{ab}(t)|{\bf y} \ra 
\nonumber\\
& = &
\left( \ba{ll} 
-i \la {\hat \varphi}_a ({\bf x}) {\hat \pi}_b({\bf y}) \ra 
& \la {\hat \varphi}_a({\bf x}) {\hat \varphi}_b({\bf y}) \ra  \\ 
\la {\hat \pi}_a({\bf x}) {\hat \pi}_b({\bf y}) \ra
&i \la {\hat \pi}_a({\bf x}) {\hat \varphi}_b({\bf y}) \ra  
\ea \right) - \frac{{\bf 1}}{2},
\end{eqnarray}
where ${\hat \varphi}=\varphi-{\bar \varphi}$ , ${\hat \pi}=\pi-{\bar \pi}$,
${\bf 1}$ is the $2N\times 2N$ unit matrix and
expectation values are calculated with the Gaussian density matrix 
$\rho[\varphi_1, \varphi_2]$. 
As in the case of zero temperature, eigenvectors of this matrix allow one to 
build mode functions. 
These functions define new creation and annihilation operators via 
Eq.(\ref{3a8}) which can be used to bring the density matrix (\ref{A1}) 
in canonical form. 

In the mean field approximation, we postulate that at each time $t$ the 
state of the system is described by a Gaussian density matrix of the 
form (\ref{A1}). 
Evolution equations in this approximation scheme are obtained by considering 
the Heisenberg evolution equations for the operators 
$\varphi\varphi$, $\varphi\pi$ and $\pi\pi$ and by calculating expectation 
values in the Gaussian state (\ref{A1}). 
As a result we find 
\begin{equation}\label{A8}
i\dot{\cal M}^{ab}(t) = [ {\cal H} , {\cal M}(t) ]^{ab} \ , 
\end{equation}
where we define ${\cal H}$ as 
\begin{eqnarray} \label{A9}
& &{\cal H} \equiv 
\left( \ba{ll} 
0 
& 1  \\ 
\Gamma
& 0  
\ea \right) \ ,  
\nonumber\\
& &\Gamma_{ab}  \equiv 
\left( -\Delta + m_0^2 
+\frac{\lambda}{6}\sum_{c=1}^{N}\bvarphi_c({\bf x})\bvarphi_c({\bf x})
+\frac{\lambda}{6}
\sum_{c=1}^{N} 
\la {\hat \varphi}_c({\bf x}){\hat \varphi}_c({\bf x}) \ra \right)\delta_{ab}
\nonumber\\
& & \qquad\qquad\qquad\qquad
+\frac{\lambda}{3}\left(
\la {\hat \varphi}_a({\bf x}){\hat \varphi}_b({\bf x}) \ra
+\bvarphi_a({\bf x}) \bvarphi_b({\bf x}) \right) \ . 
\end{eqnarray}
Thus, we can get the equations of motion governing  
the time-evolution of quantum fluctuation in thermal fluctuation 
from the above equation of motion.
Noting the explicit expression for the generalized density matrix, 
\begin{eqnarray} \label{A10}
& &{\cal M}^{ab} \nonumber\\
&=&
\left( \ba{ll} 
-2i \left[ \sqrt{G}\frac{1}{1-\xi}\sqrt{G}(\Sigma-\sigma)\right]_{ab} 
& \left[ \sqrt{G}\frac{1}{1-\xi}\sqrt{G}\right]_{ab}  \\ 
\left[\frac{1}{4}\frac{1}{\sqrt{G}}(1+\xi)\frac{1}{\sqrt{G}}+4(\Sigma+\sigma)
\sqrt{G}\frac{1}{1-\xi}\sqrt{G}(\Sigma-\sigma)\right]_{ab}
&2i \left[(\Sigma+\sigma) \sqrt{G}\frac{1}{1-\xi}\sqrt{G}\right]_{ab}  
\ea \right) \ , \nonumber\\
& & 
\end{eqnarray}
we can obtain the equations of motion for $G$, $\Sigma$, $\xi$ and $\sigma$ 
as 
\begin{eqnarray}\label{A11}
& &{\dot \Sigma}  
= \frac{1}{8}(G^{-2}-\eta^2)-2(\Sigma^2+\sigma^2)-\frac{1}{2}\Gamma \ , 
\nonumber\\
& &{\dot G}^{-1} 
= -2(G^{-1}\Sigma + \Sigma G^{-1}) + 2(\eta \sigma - \sigma \eta) \ , 
\nonumber\\
& &{\dot \eta} 
= -2(\eta\Sigma + \Sigma \eta) + 2(G^{-1} \sigma - \sigma G^{-1}) \ , 
\nonumber\\
& &{\dot \sigma} 
= -2(\sigma\Sigma + \Sigma \sigma) + \frac{1}{8}(\eta G^{-1} -G^{-1} \eta)
 \ , 
\end{eqnarray}
where we define $\eta\equiv G^{-1/2}\xi G^{-1/2}$.

\section{Equations of motion in isospin-rotating frame at finite temperature}

In this Appendix we show how to construct solutions 
which correspond to rotations of the components $\varphi_1$ and
$\varphi_2$ in isospin space with an angular frequency $\omega$. 
These solutions
are similar to those developed by Thouless and Valatin \cite{THOULESS} in
the context of mean field theory for deformed nuclei and the 
restoration of rotational symmetry when it is spontaneously broken in the
mean field ground state at zero temperature. 
Related work concerning the case of a
multicomponent scalar field can be found in references \cite{AOUISSAT,PIRNER}.

To construct these solutions we first consider the equations of motion in 
the isospin rotating frame. 
In this frame the Hamiltonian is 
\begin{equation}\label{B1}
H_{\rm rot} = H_{O(N)}-\omega_{ab} \int d^3{\bf x} 
\varphi_a({\bf x}) \pi_b({\bf x}) \ , 
\end{equation}
where we define $\omega_{ab}$ as
$\omega_{12}=-\omega_{21}\equiv \omega$ and the otherwise are equal to 0.
We can also derive the 
equation of motion for ${\cal M}$ in the isospin rotating frame : 
\begin{equation}\label{B2}
i\dot{\cal M}^{ab} = [ {\cal H}(\omega) , {\cal M}(t) ]^{ab} \ , 
\end{equation}
where we define ${\cal H}(\omega)$ as 
\begin{eqnarray} \label{B3}
& &{\cal H}(\omega) \equiv 
\left( \ba{ll} 
-\omega \tau_y 
& 1  \\ 
\Gamma
& -\omega \tau_y  
\ea \right) \ , \qquad
\tau_y \equiv
\left( \ba{llll} 
0 & -i & 0 & \cdots  \\ 
i &  0 & 0 & \cdots  \\
0 &  0 & 0 & \cdots  \\
\cdot & \cdot & \cdot & \cdots   
\ea \right) \ , 
\end{eqnarray}
and $\Gamma_{ab}$ is the same as (\ref{A9}). 
Hereafter, we consider a situation that the time-evolution of $G$, $\Sigma$ 
etc. is only originated from the rotation in the isospin space. 
In this case, the generalized density matrix ${\cal M}$ is 
``static'' in the isospin-rotating frame, namely, we obtain 
\begin{equation}\label{B4}
[ {\cal H}(\omega) , {\cal M} ] = 0 \ .
\end{equation}
Thus, ${\cal H}(\omega)$ and ${\cal M}$ are simultaneously diagonalizable. 
The equations for $Q\equiv G^{-1}$, $\Sigma$, $\eta \ (\xi)$ and $\sigma$ are 
obtained by replacing ${\dot Q}$ into $i\omega [ Q, \tau_y]$ in 
Eq.(\ref{A11}).  
For the condensate or mean field, we obtain the following equation in the 
isospin-rotating frame : 
\begin{equation}\label{B5}
\omega^2 \bvarphi_a =  
\left\{
\left( -\Delta + m_0^2 + \frac{\lambda}{6} \bvarphi^2 +  \frac{\lambda}{6} 
 \la {\hat \varphi}_c({\bf x}){\hat \varphi}_c({\bf x})\ra
 \right) \delta_{ab} 
+ \frac{\lambda}{3} \la {\hat \varphi}_a({\bf x}){\hat \varphi}_b({\bf x})\ra
\right\} \bvarphi_b.
\end{equation}

\section{Spectral decomposition of generalized density matrix}

In this Appendix, we give a spectral decomposition for the generalized 
density matrix ${\cal M}$ in the $\sigma=0$ case in terms of mode functions, 
namely, the solution of the Hartree-Bogoliubov equations.

We will start with a situation that the condensate or mean field 
${\bvarphi}$ does not depend on space and time in the isospin-rotating 
frame, pointing in the direction of the first isospin :
\begin{equation}\label{C1}
\bvarphi = 
\left( \ba{l} 
\varphi_0  \\ 
0 \\
\cdot \\
\cdot \\
0  
\ea \right) 
= \varphi_0 \ \delta_{a1} \ . 
\end{equation}
As is shown later the averaged values 
$\la {\hat \varphi}_a({\bf x}){\hat \varphi}_b({\bf x}) \ra$
have a diagonal form in the isospins space. 
The mass matrix 
$M_{ab}^2$ from $\Gamma$ in Eq.(\ref{A9}) reads 
\begin{eqnarray} \label{C2}
\Gamma_{ab} &=& \Delta \ \delta_{ab} + M_{ab}^2 \ , \nonumber\\
M_{ab}^2 &\equiv& \left(m_0^2 + \frac{\lambda}{6}
\sum_{c=1}^{N}(\bvarphi_c({\bf x}))^2 + \frac{\lambda}{6}
\sum_{c=1}^{N}\la{\hat \varphi}_c({\bf x}){\hat \varphi}_c({\bf x})\ra 
\right) \delta_{ab}+\frac{\lambda}{3}
(\la{\hat \varphi}_a({\bf x}){\hat \varphi}_b({\bf x})\ra 
+\bvarphi_a \bvarphi_b) \nonumber\\
&=& M_a^2 \delta_{ab} \ , \nonumber\\
M_a^2 &\equiv&
m_0^2 + \frac{\lambda}{6}
\varphi_0^2 + \frac{\lambda}{6}
\sum_{c=1}^{N}\la{\hat \varphi}_c({\bf x}){\hat \varphi}_c({\bf x})\ra 
+\frac{\lambda}{3}
(\la{\hat \varphi}_a({\bf x}){\hat \varphi}_a({\bf x})\ra 
+\varphi_0^2 \delta_{a1}) \ . 
\end{eqnarray}
Further, we will show later that 
$\la {\hat \varphi}_c({\bf x}) {\hat \varphi}_c({\bf x})\ra$ 
for $c \ge 3$ have same values because isospin rotation occurs only 
in $c=1$- and 2-components.
Thus, we can set up the masses as 
\begin{eqnarray} \label{C3}
& & M_1^2 \equiv M_0^2 + \mu_0^2 \ , \nonumber\\
& & M_2^2 \equiv M_0^2 - \mu_0^2 \ , \nonumber\\
& & M_3^2 = M_4^2 = \cdots = M_N^2 \equiv \mu^2 \ , 
\end{eqnarray}
where we define $M_0^2$ and $\mu_0^2$ as 
\begin{eqnarray} \label{C4}
& &M_0^2 \equiv m_0^2+\frac{\lambda}{3}\varphi_0^2+
\frac{\lambda}{3}(\la{\hat \varphi}_1({\bf x}){\hat \varphi}_1({\bf x})\ra 
+\la{\hat \varphi}_2({\bf x}){\hat \varphi}_2({\bf x})\ra)
+\frac{\lambda}{6}\sum_{c=3}^{N}
\la{\hat \varphi}_c({\bf x}){\hat \varphi}_c({\bf x})\ra \ , \nonumber\\
& &\mu_0^2 \equiv \frac{\lambda}{6}\varphi_0^2+
\frac{\lambda}{6}(\la{\hat \varphi}_1({\bf x}){\hat \varphi}_1({\bf x})\ra 
-\la{\hat \varphi}_2({\bf x}){\hat \varphi}_2({\bf x})\ra)
\ .
\end{eqnarray}
Then, the matrix $\Gamma$ has a simple form : 
\begin{equation} \label{C5}
\Gamma_{ab} = (-\Delta + m_a^2 + \mu_0^2 \tau_z) \delta_{ab} \ , 
\end{equation}
where $(\tau_z)_{11}=-(\tau_z)_{22}=1$ and the others are equal to 0. 
Here we define 
$m_1^2=m_2^2\equiv M_0^2$ and $m_a^2 \equiv \mu^2$ for $a \ge 3$.
Using the above simple expressions, we can easily get the eigenvalues 
for ${\cal H}(\omega)$ in momentum representation from the following 
eigenvalue equations : 
\begin{equation} \label{C6}
\left( \ba{ll} 
-\omega \tau_y 
& 1 \\ 
\Gamma({\bf k})
& -\omega \tau_y   
\ea \right)
\left( \ba{l}
u_{\bf k} \\
v_{\bf k}
\ea \right)
= E({\bf k})
\left( \ba{l}
u_{\bf k} \\
v_{\bf k}
\ea \right) \ ,
\end{equation}
where $(u_{\bf k}, v_{\bf k})$ and $E({\bf k})$ are eigenvectors and 
eigenvalues for ${\cal H}(\omega)$ in momentum representation, respectively.
As a result, we get 
\begin{eqnarray} \label{C7}
& &E^2({\bf k}, a=\pm)= {\bf k}^2+M_0^2+\omega^2\pm
\sqrt{\mu_0^4+4\omega^2({\bf k}^2+M_0^2)} \ , \nonumber\\
& &E^2({\bf k}, a\ge 3)={\bf k}^2+\mu^2 \ , 
\end{eqnarray}
where $+$, $(-)$ means $a=1$, (2).

In order to get the spectral decomposition for the generalized density 
matrix ${\cal M}$ in terms of $(u_{\bf k}, v_{\bf k})$, 
it should be noted that the matrix ${\cal M}$ is expressed with the help of  
three matrices as 
\begin{equation} \label{C8}
{\cal M}=V_{\sigma}
\left( \ba{ll}
\frac{1}{2}\sqrt{\frac{1+\xi}{1-\xi}} 
& 0 \\
0 
& -\frac{1}{2}\sqrt{\frac{1+\xi}{1-\xi}} 
\ea \right) 
W_{\sigma} \ , 
\end{equation}
where 
\begin{eqnarray} \label{C9}
& &V_{\sigma}=
\left( \ba{ll}
G^{\frac{1}{2}} 
& G^{\frac{1}{2}} \\
2i(\Sigma+\sigma)G^{\frac{1}{2}}+G^{-\frac{1}{2}}\frac{\sqrt{1-\xi^2}}{2} 
& 2i(\Sigma+\sigma)G^{\frac{1}{2}}-G^{-\frac{1}{2}}\frac{\sqrt{1-\xi^2}}{2} 
\ea \right) \ , \nonumber\\
& &W_{\sigma}=
\left( \ba{ll}
\frac{1}{2}G^{-\frac{1}{2}}-2i\frac{1}{\sqrt{1-\xi^2}}G^{\frac{1}{2}}
(\Sigma-\sigma) 
& \frac{1}{\sqrt{1-\xi^2}}G^{\frac{1}{2}} \\
\frac{1}{2}G^{-\frac{1}{2}}+2i\frac{1}{\sqrt{1-\xi^2}}G^{\frac{1}{2}}
(\Sigma-\sigma)
& -\frac{1}{\sqrt{1-\xi^2}}G^{\frac{1}{2}} 
\ea \right) \ .
\end{eqnarray}

We will now look for solutions in which 
the imaginary part of the mixing parameter,
$\sigma$, is equal to 0. 
This corresponds to 
the ``static'' case in the 
rotating frame. Then, the matrix ${\cal M}$ is simply expressed as 
\begin{equation} \label{C10}
{\cal M}=V
\left( \ba{ll}
\frac{1}{2}\sqrt{\frac{1+\xi}{1-\xi}} 
& 0 \\
0 
& -\frac{1}{2}\sqrt{\frac{1+\xi}{1-\xi}} 
\ea \right) 
V^{-1} \ , 
\end{equation}
where $V\equiv V_{\sigma=0}$ and $V^{-1} (=W_{\sigma=0})$ is 
an inverse operator matrix of $V$. 
Then, for positive eigenvalues of ${\cal M}$, we obtain 
\begin{equation} \label{C12}
{\cal M} \pmatrix{ |u_c({\bf k})\ra \cr |v_c({\bf k})\ra}
=\frac{1}{2}\sqrt{\frac{1+\xi_c({\bf k})}{1-\xi_c({\bf k})}}
\pmatrix{ |u_c({\bf k})\ra \cr |v_c({\bf k})\ra} \ , 
\qquad
\pmatrix{\frac{1}{n_c}|u_c({\bf k})\ra \cr \frac{1}{n_c}|v_c({\bf k})\ra}
\equiv V
\pmatrix{|w_c({\bf k})\ra \cr 0} \ ,
\end{equation}
where $\xi_c({\bf k})$ and $|w_c({\bf k})\ra$ are eigenvalue and eigenvector 
for the operator $\xi(t)$, respectively. Here, $n_c$ are normalization 
factors. It should be noted that the $|v_c({\bf k})\ra$ is related to 
$|u_c({\bf k})\ra$ as follows : 
\begin{equation} \label{C13}
|v_c({\bf k})\ra = \left(\frac{1}{2\sqrt{G}}\sqrt{1-\xi^2}
\frac{1}{\sqrt{G}}+2i\Sigma\right) |u_c({\bf k})\ra \ , 
\end{equation}
where 
$|u_c({\bf k})\ra \equiv n_c G^{\frac{1}{2}} |w_c({\bf k})\ra$. 
From (\ref{C12}), 
we obtain the spectral decomposition for the generalized density matrix 
${\cal M}$ : 
\begin{eqnarray}\label{C14}
& &{\cal M}=
\sum_{{\bf k}, (E>0)}\sum_{c=1}^{N}
\biggl\{
\pmatrix{|u_c({\bf k})\ra \cr |v_c({\bf k})\ra}
\frac{1}{2}\sqrt{\frac{1+\xi_c({\bf k})}{1-\xi_c({\bf k})}}
(\la v_c({\bf k})| , \la u_c({\bf k})| ) \nonumber\\
& &\qquad\qquad\qquad
+
\pmatrix{|u_c^*({\bf k})\ra \cr -|v_c^*({\bf k})\ra}
\frac{1}{2}\sqrt{\frac{1+\xi_c({\bf k})}{1-\xi_c({\bf k})}}
(-\la v_c^*({\bf k})| , \la u_c^*({\bf k})| ) \biggl\} \ ,
\end{eqnarray}
where we used the relations 
\begin{equation} \label{C15}
{\cal M} \pmatrix{|u_c^*({\bf k})\ra \cr -|v_c^*({\bf k})\ra }
=-\frac{1}{2}\sqrt{\frac{1+\xi_c({\bf k})}{1-\xi_c({\bf k})}}
 \pmatrix{|u_c^*({\bf k})\ra \cr -|v_c^*({\bf k})\ra }
\end{equation}
for the eigenvectors of negative eigenvalues.
The notation $E>0$ means that the sum runs over the eigenvectors 
with positive eigenvalues. 
It should be noted that ${\cal H}(\omega)$ and ${\cal M}$ are simultaneously 
diagonalizable because the relation $[{\cal H}(\omega) , {\cal M} ]=0$ 
is satisfied in the rotating frame.
Applying adequate normalization factors $n_c$, we get the ortho-normalization 
and completeness relations as 
\begin{equation} \label{C16}
\int d{\bf x} \{v_c({\bf k},{\bf x})^* u_c({\bf k}', {\bf x})
+u_c({\bf k}, {\bf x})^* v_c({\bf k}', {\bf x}) \} 
= \pm \delta_{{\bf k}{\bf k}'} \ 
\end{equation}
with $+$ $(-)$ sign for positive (negative) 
eigenvalues, and
\begin{equation} \label{C17}
\sum_{{\bf k}, (E>0)}\sum_{c}
\biggl\{ \pmatrix{|u_c({\bf k})\ra \cr |v_c({\bf k})\ra }
(\la v_c({\bf k})|, \la u_c({\bf k}) |)
- \pmatrix{|u_c^*({\bf k})\ra \cr -|v_c^*({\bf k})\ra }
(-\la v_c^*({\bf k})|, \la u_c^*({\bf k}) |)
\biggl\}
=1 \ .
\end{equation}

Finally, it is necessary to determine $\xi_c({\bf k})$ in ${\cal M}$, 
while the eigenvectors are directly calculable.
Since we treat the case $\sigma$ (the imaginary part of the mixing parameter) 
is equal to 0, the generalized density matrix has a simple form in Eq. 
(\ref{C10}). Thus, we can write the entropy from this density matrix 
in the unit $k_{B}=1$ as 
\begin{equation} \label{C18}
S=\sum_{{\bf k}, c} [ (1+n_{{\bf k},c}) \ln (1+n_{{\bf k},c})
-n_{{\bf k},c} \ln n_{{\bf k},c} ] \ , 
\end{equation}
where the $n_{{\bf k},c}$'s are defined by
\begin{equation} \label{C19}
n_{{\bf k},c} 
= \frac{1}{2} \left(\sqrt{\frac{1+\xi_c({\bf k})}{1-\xi_c({\bf k})}}-1
\right) \ .
\end{equation}
Thus, $\xi_c({\bf k})$ or $n_{{\bf k},c}$ occurring in the generalized density 
matrix can be obtained by imposing that the free energy $F(\omega)$ in the 
rotating frame 
\begin{equation} \label{C20}
F(\omega) = \la {\cal H}(\omega) \ra - \frac{S}{\beta}
\end{equation}
is stationary because all variables are time-independent in the rotating 
frame.
Here, $\beta\equiv 1/T$ is inverse of temperature. 
The variation of $\la {\cal H}(\omega) \ra$ with respect to 
$n_{{\bf k},c}$ is easily obtained by noting that the change 
$\delta {\cal M}$ is given by 
\begin{equation} \label{C21}
\delta{\cal M} = \sum_{{\bf k},c} 
\pmatrix{ u_c({\bf k}) \cr v_c({\bf k})} \delta n_{{\bf k},c}
(u_c({\bf k})^*, v_c({\bf k})^*) . 
\end{equation}
Since mode functions 
$(u_c({\bf k}), v_c({\bf k}))$ are eigenstates of the matrix 
${\cal H}(\omega)$ due to\break
$[{\cal H}(\omega) , {\cal M} ]=0$, we have 
\begin{equation} \label{C22}
\delta \la {\cal H}(\omega) \ra = {\rm Trace}({\cal H}(\omega)\delta{\cal M})
=\sum_{{\bf k},c}E_{{\bf k},c} \delta n_{{\bf k},c} \ .
\end{equation}
Moreover, since we obtain 
$\delta S = \sum_{{\bf k},c} \delta n_{{\bf k},c} 
\ln ((1+n_{{\bf k},c})/n_{{\bf k},c})$, 
thus $\delta F(\omega) = 0$ gives us 
\begin{equation} \label{C23}
n_{{\bf k},c} = \frac{1}{{\rm e}^{\beta E_{{\bf k},c}}-1} \ , 
\quad \hbox{\rm or} \quad
\xi_c({\bf k}) = \frac{1}{\cosh \beta E_{{\bf k},c}} \ .
\end{equation}
One can understand that $n_{{\bf k},c}$ are nothing but boson distribution 
functions.

\section{Gap equation in the rotating frame}

In this Appendix, we give solutions for 
$\la {\hat \varphi}_a({\bf x}) {\hat \varphi}_b ({\bf x}) \ra$ 
in terms of the mode functions.
In a first step we consider the special case of a Gaussian density matrix 
with zero momentum.
To treat the most general case of a Gaussian density matrix with finite 
momentum we
exploit in a second step the evolution equations 
for mode functions to generate in a natural way solutions with a finite
value of the momentum. 
We can get the results with the finite momentum by following the method 
developed in \S 5.
The gap equation to determine the chiral order parameter 
$\varphi_0$ is derived in the closed set of equations.

From the spectral decomposition of ${\cal M}$, we can write 
$\la {\hat \varphi}_a({\bf x}) {\hat \varphi}_b ({\bf x}) \ra$ 
explicitly as 
\begin{eqnarray} \label{D1}
\la {\hat \varphi}_a({\bf x}) {\hat \varphi}_b ({\bf x}) \ra 
&=& {\cal M}_{12}^{ab}({\bf x}, {\bf x}) \nonumber\\
&=& \sum_{{\bf k},(E>0)} \sum_{\sigma}
\frac{1}{2}\sqrt{\frac{1+\xi_\sigma(E_{{\bf k},\sigma})}
{1-\xi_\sigma(E_{{\bf k},\sigma})}} \nonumber\\
& & \qquad\qquad \times
[u_{\sigma{\bf k}}^a({\bf x})u_{\sigma{\bf k}}^{b*}({\bf x})
+u_{\sigma{\bf k}}^{a*}({\bf x}) u_{\sigma{\bf k}}^b({\bf x})] \ . 
\end{eqnarray}
We can obtain 
$\la {\hat \varphi}_a({\bf x}) {\hat \varphi}_b ({\bf x}) \ra$ 
by calculating directly the mode functions 
$u_{\sigma{\bf k}}^a({\bf x})$ for ${\cal H}(\omega)$ such as 
\begin{equation} \label{D2}
\la {\hat \varphi}_a({\bf x}) {\hat \varphi}_b ({\bf x}) \ra
=\delta_{ab} \sum_{\bf k} \frac{1}{\sqrt{{\bf k}^2+\mu^2}}\left(
n_{{\bf k},a}+\frac{1}{2} \right) \qquad ({\rm for}\ a,b=3,4,\cdots, N) \ ,
\end{equation}
where $\mu^2$ is defined by Eqs. (\ref{C2}) and (\ref{C3}). 
For $a\ge 3$, we can easily obtain the above expression by calculating 
the mode functions directly. 
Although the isospin-rotating components, 
$\la {\hat \varphi}_a({\bf x}) {\hat \varphi}_b ({\bf x}) \ra$ 
($a, b = 1,2$), are easily calculable through mode functions, we give 
another derivation here.
The two methods, of course, give identical results. 

Let us introduce the following $N \times N$ matrix operator :
\begin{eqnarray} \label{D3}
\la {\bf x}, x_0|S_{ab}|{\bf y}, y_0\ra 
&=& \theta (x_0-y_0) \sum_{{\bf k}, (E>0)}\sum_{\sigma}
\sqrt{\frac{1+\xi_{\sigma}({\bf k})}{1-\xi_{\sigma}({\bf k})}}
u_{\sigma{\bf k}}^a({\bf x},x_0) u_{\sigma{\bf k}}^{b*}({\bf y},y_0) 
\nonumber\\
& &+ \theta (y_0-x_0) \sum_{{\bf k}, (E>0)}\sum_{\sigma}
\sqrt{\frac{1+\xi_{\sigma}({\bf k})}{1-\xi_{\sigma}({\bf k})}}
u_{\sigma{\bf k}}^{a*}({\bf x},x_0) u_{\sigma{\bf k}}^b({\bf y},y_0)
\nonumber\\
&\equiv& \la {\bf x}|S_{ab}(x_0, y_0)|{\bf y}\ra \ , 
\end{eqnarray}
where the mode functions $u_{\sigma{\bf k}}^a({\bf x}, x_0)$, together with 
$v_{\sigma{\bf k}}^a({\bf x}, x_0)$,  
satisfy the following equations of motion in the isospin-rotating frame :
\begin{equation} \label{D4}
i\partial_{x_0}
\pmatrix{u_{\sigma{\bf k}}({\bf x},x_0) \cr
v_{\sigma{\bf k}}({\bf x}, x_0)} 
=
\left( \ba{ll} 
-\omega \tau_y 
& 1 \\ 
\Gamma({\bf k})
& -\omega \tau_y   
\ea \right)
\left( \ba{l}
u_{\sigma{\bf k}}({\bf x}, x_0) \\
v_{\sigma{\bf k}}({\bf x}, x_0)
\ea \right)
 \ ,
\end{equation}
where $\Gamma$ has been already defined in Eq.(\ref{C5}).
Eliminating $v_{\sigma{\bf k}}({\bf x},x_0)$, we get the following 
Klein-Gordon-type equations of motion :
\begin{equation} \label{D5}
(-\partial_{x_0}^2 - \Gamma+\omega^2+2i\omega \tau_y \partial_{x_0})
u_{\sigma{\bf k}}({\bf x},x_0)=0 \ .
\end{equation}
The matrix operator $S$ is related to 
$\la {\hat \varphi}_a({\bf x}) {\hat \varphi}_b ({\bf x}) \ra$ 
through the relation 
\begin{equation} \label{D6}
\la {\bf x}, x_0 | S | {\bf x}, x_0 \ra = {\cal M}_{12}({\bf x}, {\bf x}) \ 
\end{equation}
and Eq.(\ref{D1}). 
Here we note that $u_{\sigma{\bf k}}({\bf x}, x_0)$ are also eigenvectors 
for the operator $\xi(x_0)$ from the relation 
$|u_c \ra \propto G^{1/2} |w_c \ra$
because $\xi(x_0)$ 
and $G(x_0)$ commute due to the equation of motion for 
the imaginary part of the mixing parameter, $\sigma$, in the case $\sigma=0$. 
We introduce the Feynman propagator 
\begin{eqnarray} \label{D7}
\la {\bf x}, x_0|{\bar S}_{ab}|{\bf y}, y_0\ra 
&\equiv& 
\la {\bf x}|\left(\frac{1+\xi(x_0)}{1-\xi(x_0)}\right)^{-\frac{1}{4}}
S_{ab}(x_0, y_0)
\left(\frac{1+\xi(y_0)}{1-\xi(y_0)}\right)^{-\frac{1}{4}} 
|{\bf y} \ra \nonumber\\
&=& 
\theta (x_0-y_0) \sum_{{\bf k}, (E>0)}\sum_{\sigma}
u_{\sigma{\bf k}}^a({\bf x},x_0) u_{\sigma{\bf k}}^{b*}({\bf y},y_0)
\nonumber\\
& &+ \theta (y_0-x_0) \sum_{{\bf k}, (E>0)}\sum_{\sigma}
u_{\sigma{\bf k}}^{a*}({\bf x},x_0) u_{\sigma{\bf k}}^b({\bf y},y_0)
\ , 
\end{eqnarray}
which satisfies 
\begin{equation} \label{D8}
(\partial_{x_0}^2 + \Gamma-\omega^2-2i\omega \tau_y \partial_{x_0})
\la {\bf x}, x_0 | {\bar S} | {\bf y}, y_0 \ra 
= -i \delta^4(x-y)
\end{equation}
by using the equations of motion for $u_{\sigma{\bf k}}({\bf x},x_0)$ 
and $v_{\sigma{\bf k}}({\bf x}, x_0)$ and the completeness relation. 
Thus, it is symbolically written by 
\begin{equation} \label{D9} 
\la {\bf x}, x_0 | {\bar S} | {\bf y}, y_0 \ra 
=
\frac{-i}{\partial_{x_0}^2 + \Gamma - \omega^2 -2i\omega \tau_y \partial_{x_0} 
-i\epsilon} 
\end{equation}
with infinitesimal positive constant $\epsilon$. 
From the Fourier-transformation for the propagator 
$\la {\bf x}, x_0|{\bar S}|{\bf y},y_0 \ra$, namely 
\begin{equation} \label{D10}
{\bar S}(x, y)=\int \frac{d^4 p}{(2\pi)^4} {\bar S}(p) 
{\rm e}^{-ip(x-y)} \ , 
\end{equation}
we obtain
\begin{eqnarray} \label{D11}
{\bar S}(p) &=&
\left(\frac{1+\xi(p_0)}{1-\xi(p_0)}\right)^{-\frac{1}{4}}
S(p)
\left(\frac{1+\xi(p_0)}{1-\xi(p_0)}\right)^{-\frac{1}{4}}\nonumber\\
&=&
\frac{i}{p_0^2-{\bf p}^2-m^2-\mu_0^2 \tau_z +\omega^2-2\omega p_0 \tau_y 
+i\epsilon} \ .
\end{eqnarray}
In this way, we get the Fourier transformation of 
$\la {\bf x}, x_0|S|{\bf y},y_0 \ra$ :
\begin{equation} \label{D12}
S(p) =
(2n(p_0)+1) 
\frac{i}{(p_0-\omega\tau_y)^2-{\bf p}^2-m^2-\mu_0^2 \tau_z +i\epsilon} \ ,
\end{equation}
where
\begin{equation} \label{D13}
n(p_0)=\frac{1}{{\rm e}^{\beta p_0}-1}
\end{equation}
with the help of Eqs.(\ref{C19}) and (\ref{C23}).

Let us exploit in the second step the evolution equations for mode functions 
to generate solutions with a finite value of the three-momentum ${\bf q}$ 
in addition to $\omega$.
This corresponds to using the mode function 
$u_{\sigma {\bf k}}^{\bf q}({\bf x}, x_0)
=e^{i{\bf q}\cdot {\bf x}\tau_y}
u_{\sigma {\bf k}} ({\bf x}, x_0)$. 
We obtain the following general expression :
\begin{equation} \label{D14}
S(p) =
(2n(p_0)+1) 
\frac{i}{(p_0-\omega\tau_y)^2-({\bf p}-{\bf q}\tau_y)^2
-m^2-\mu_0^2 \tau_z +i\epsilon} \ .
\end{equation}
We can get ${\cal M}_{12}^{ab}({\bf x}, {\bf x})$  
or ${\cal M}_{12}^{ab}({\bf k})$, which are introduced by 
${\cal M}_{12}^{ab}({\bf x}, {\bf x})
=\sum_{\bf k}{\cal M}_{12}^{ab}({\bf k})$, performing the inverse Fourier 
transformation from $S(p)$ by the use of the relation of Eq.(\ref{D6}) :
\begin{equation} \label{D15}
{\cal M}_{12}({\bf k})
=-\int\frac{d k_0}{2\pi i} (2n(k_0)+1)
\frac{1}{(k_0-\omega\tau_y)^2-({\bf k}-{\bf q}\tau_y)^2
-m^2-\mu_0^2\tau_z +i\epsilon} \ .
\end{equation}

Let us consider the propagator in the subspace of the first two isospins. 
First, we concentrate on the time-like $q^2$ as 
$q^2=\omega^2$ with ${\bf q}={\bf 0}$. 
After integration with respect to $k_0$, we have 
${\cal M}_{12}^{ab}({\bf k})$ for $a, b=1,2$ : 
\begin{eqnarray} \label{D16}
\la {\hat \varphi}_1({\bf x}) {\hat \varphi}_1({\bf x}) \ra
&=&{\cal M}_{12}^{11}({\bf x}, {\bf x}) 
=S_{11}(x, x) \nonumber\\
&=&
\int\frac{d^3{\bf k}}{(2\pi)^3} 
\biggl\{
(2n(E_+)+1)\left[\frac{1}{4E_+}+\frac{2\omega^2+\mu_0^2}{2E_+(E_+^2-E_-^2)}
\right] \nonumber\\
& &\qquad\qquad
+(2n(E_-)+1)\left[\frac{1}{4E_-}-\frac{2\omega^2+\mu_0^2}{2E_-(E_+^2-E_-^2)}
\right]\biggl\} \ , \nonumber\\
\la {\hat \varphi}_2({\bf x}) {\hat \varphi}_2({\bf x}) \ra
&=&{\cal M}_{12}^{22}({\bf x}, {\bf x}) 
=S_{22}(x, x) \nonumber\\
&=&
\int\frac{d^3{\bf k}}{(2\pi)^3} 
\biggl\{
(2n(E_+)+1)\left[\frac{1}{4E_+}+\frac{2\omega^2-\mu_0^2}{2E_+(E_+^2-E_-^2)}
\right] \nonumber\\
& &\qquad\qquad
+(2n(E_-)+1)\left[\frac{1}{4E_-}-\frac{2\omega^2-\mu_0^2}{2E_-(E_+^2-E_-^2)}
\right]\biggl\} \ , \nonumber\\
\la {\hat \varphi}_a({\bf x}) {\hat \varphi}_b({\bf x}) \ra
&=& 0 \qquad\qquad {\rm for} \ a \neq b \ , 
\end{eqnarray}
where we denote 
$\la {\bf x}, x_0 | S | {\bf y}, y_0 \ra$ as $S(x, y)$. 
Here, $E_{\pm}$ are defined in (\ref{C7}) and 
\begin{equation} \label{D17}
n(E)=\frac{1}{{\rm e}^{\beta E}-1} \ .
\end{equation}
Also, $\la {\hat \varphi}_c({\bf x}) {\hat \varphi}_c({\bf x}) \ra$ 
for $c\ge 3$ are given in Eq.(\ref{D2}).  
Thus, our assumption that $M_{ab}^2$ has a diagonal form is valid 
because 
$\la {\hat \varphi}_a({\bf x}) {\hat \varphi}_b({\bf x}) \ra = 0$ 
for $a\neq b$.

Secondly, let us consider the space-like $q^2$ as $q^2=-{\bf q}^2$ 
with $\omega=0$. 
In this case, we can get ${\cal M}_{12}^{ab}({\bf x}, {\bf x})$ 
in the same way as the time-like case : 
\begin{eqnarray} \label{D18}
\la {\hat \varphi}_1({\bf x}) {\hat \varphi}_1({\bf x}) \ra
&=&{\cal M}_{12}^{11}({\bf x}, {\bf x}) 
=S_{11}(x, x) \nonumber\\
&=&
\int\frac{d^3{\bf k}}{(2\pi)^3} 
\biggl\{
(2n(E_+)+1)\left[\frac{1}{4E_+}+\frac{\mu_0^2}{2E_+(E_+^2-E_-^2)}
\right] \nonumber\\
& &\qquad\qquad
+(2n(E_-)+1)\left[\frac{1}{4E_-}-\frac{\mu_0^2}{2E_-(E_+^2-E_-^2)}
\right]\biggl\} \ , \nonumber\\
\la {\hat \varphi}_2({\bf x}) {\hat \varphi}_2({\bf x}) \ra
&=&{\cal M}_{12}^{22}({\bf x}, {\bf x}) 
=S_{22}(x, x)\nonumber\\
&=&
\int\frac{d^3{\bf k}}{(2\pi)^3} 
\biggl\{
(2n(E_+)+1)\left[\frac{1}{4E_+}-\frac{\mu_0^2}{2E_+(E_+^2-E_-^2)}
\right] \nonumber\\
& &\qquad\qquad
+(2n(E_-)+1)\left[\frac{1}{4E_-}+\frac{\mu_0^2}{2E_-(E_+^2-E_-^2)}
\right]\biggl\} \ , \nonumber\\
\la {\hat \varphi}_a({\bf x}) {\hat \varphi}_b({\bf x}) \ra
&=& 0 \qquad\qquad {\rm for} \ a \neq b \ . 
\end{eqnarray}
Here, $E_{\pm}\ (>0)$ are obtained by the poles of the integrand 
in (\ref{D15}) 
with $\omega=0$ :
\begin{equation}\label{D19}
E_{\pm}^2({\bf k})
={\bf k}^2+{\bf q}^2+M_0^2 \pm
\sqrt{\mu_0^4+4({\bf k}\cdot {\bf q})^2} \ .
\end{equation}

Finally, the gap equation in the isospin-rotating frame 
at finite temperature is derived from the equation of motion for $\bvarphi$ 
with $\bvarphi_a = \varphi_0 \delta_{a1}$. 
We obtain the gap equation for $\varphi_0$ in the symmetry broken phase :
\begin{equation} \label{D20}
\frac{\lambda}{6}\varphi_0^2
=\omega^2-{\bf q}^2-m_0^2-\frac{\lambda}{6}\sum_{c=1}^{N}
\la {\hat \varphi}_c({\bf x}) {\hat \varphi}_c({\bf x}) \ra
-\frac{\lambda}{3}
\la {\hat \varphi}_1({\bf x}) {\hat \varphi}_1({\bf x}) \ra
\end{equation}
due to 
$\la {\hat \varphi}_a({\bf x}) {\hat \varphi}_b({\bf x}) \ra=0$ 
for $a\neq b$. 
By using the mass $M_1^2$, the above gap equation is recast into simpler 
form as 
\begin{equation} \label{D21}
\frac{\lambda}{3}\varphi_0^2 = -\omega^2+{\bf q}^2+M_1^2 \ .
\end{equation}







\begin{thebibliography}{30}
\bibitem{BLAIZOT} J. P. Blaizot and A. Krzywicki, \prb{D46} (1992) 246 
\bibitem{WILCZEK} K. Rajagopal and F. Wilczek, \npb{B404} (1993) 577 
\bibitem{HOLMAN} D. Boyanovsy, H. J. de Vega and R. Holman, \prb{D51}
(1995) 734
\bibitem{SINGH} D. Boyanovsky, H. J. de Vega, R. Holman, D. S. Lee and A.
Singh, \prb{D51} (1995) 4419
\bibitem{ANSELM}  A. A. Anselm, M. G. Ryskin, Phys. Lett. {\bf B266} (1991)
482; A. A. Anselm, M. G. Ryskin and A. G. Shuvaev, Z. Phys. {\bf
A354} (1996) 333 
\bibitem{COOPER} F. Cooper, Y. Kluger, E. Mottola, and J. Paz, Phys. Rev. {\bf
D51} (1995) 2377 
\bibitem{GUTH} A. Guth and S. Y. Pi, Phys. Rev. {\bf D32} (1985) 1899
\bibitem{JACKIW} O. Eboli, R. Jackiw and S. Y. Pi, Phys. Rev. {\bf D37}
(1988) 3557
\bibitem{DEVEGA} D. Boyanovsky, H. J. de Vega and R. Holman, Phys. Rev. {\bf
D49} (1994) 2769
\bibitem{HECTOR} D. Boyanovsky, M. D'Attanasio, H. J. de Vega 
and R. Holman,  Phys. Rev. {\bf D54} (1996) 1748
\bibitem{SAMIULLAH} S. Y. Pi and M. Samiullah, Phys. Rev. {\bf D36} (1987)
3128 
\bibitem{MOTTOLA} F. Cooper and E. Mottola, Phys. Rev. {\bf D36} (1987) 3138
\bibitem{ENIKOVA} M. M. Enikova, V. I. Karloukovski and C. I. Vlechev, Nucl.
Phys. {\bf B151} (1979) 172
\bibitem{PEIERLS} R. E. Peierls, Surprises in Theoretical Physics, Princeton
University Press, Princeton, 1979, section 1.3 
\bibitem{PHYREV96} D. Vautherin and T. Matsui, Phys. Rev. {\bf D55} 
(1996) 4492
\bibitem{PHYLET98} D. Vautherin and T. Matsui, Phys. Lett. {\bf B437} 
(1988) 173
\bibitem{LIN} A. K. Kerman and C. Y. Lin, Ann. Phys. {\bf 241} (1995) 185
\bibitem{CORNWALL} J. Cornwall, R. Jackiw and E. Tomboulis, Phys. Rev. {\bf
D10} (1974) 2428
\bibitem{BJORKEN} See e.g., J. D. Bjorken and S. Drell, Relativistic Quantum
Mechanics, McGraw- Hill, New York, 1964
\bibitem{BAYM} G. Baym and G. Grinstein, Phys. Rev. {\bf D15} (1977) 2897 
\bibitem{ROH} H.-S. Roh and T. Matsui, Eur. Phys. J. A {\bf 1} (1998) 205
\bibitem{THOULESS} D. J. Thouless and J. G. Valatin, Nucl. Phys. {\bf 31}
(1962) 211
\bibitem{AOUISSAT} Z. Aouissat, G. Chanfray, P. Schuck and J. Wambach, Nucl.
Phys. {\bf A603} (1996) 458
\bibitem{PIRNER} H. W. L. Naus, T. Gasenzer, H. J. Pirner, Ann. Physik 6 
(1997) 287 
\bibitem{KERMAN} A. K. Kerman, C. Martin and D. Vautherin,
\prb{D47} (1993) 632
\bibitem{TSUE} Y. Tsue and Y. Fujiwara, Prog. Theor. Phys. {\bf 86} (1991)
443 and {\it ibid.} 469
\bibitem{ZINN} J. Zinn Justin, Field theory and critical phenomena
\bibitem{BEG} M. A. B. B\'eg and R. C. Furlong, Phys. Rev. {\bf D31} (1985)
1370
\bibitem{MARTIN} C. Martin, Phys. Rev. {\bf D52} (1995) 7121
\end{thebibliography}
\end{document}